\documentclass[twocolumn]{aastex631}

\usepackage{txfonts}
\usepackage{graphicx}
\usepackage{color}

\definecolor{applegreen}{rgb}{0.55, 0.71, 0.0}
\definecolor{dr}{rgb}{0.5, 0.0, 0.0}
\definecolor{dg}{rgb}{0.0, 0.5, 0.0}
\definecolor{db}{rgb}{0.0, 0.3, 0.8}
\definecolor{orange}{rgb}{0.9, 0.5, 0.0}

\begin{document}
\title{Filament Leg--Leg Reconnection as a Source of Prominent Supra-Arcade Downflows}

\correspondingauthor{Jaroslav Dud\'{i}k}
\email{jaroslav.dudik@asu.cas.cz}

\author[0000-0003-1308-7427]{Jaroslav Dud\'{i}k}
\affil{Astronomical Institute, Academy of Sciences of the Czech Republic, Fri\v{c}ova 298, 25165 Ond\v{r}ejov, Czech Republic}

\author[0000-0001-5810-1566]{Guillaume Aulanier}
\affil{Sorbonne Universit\'e, Observatoire de Paris - PSL, \'Ecole Polytechnique, Institut Polytechnique de Paris, CNRS, Laboratoire de physique des plasmas (LPP), 4 place Jussieu, F-75005 Paris, France}
\affil{Rosseland Centre for Solar Physics, Institute for Theoretical Astrophysics, Universitetet i Oslo, P.O. Box 1029, Blindern, 0315 Oslo, Norway}

\author[0000-0001-9559-4136]{Jana Ka\v{s}parov\'{a}}
\affil{Astronomical Institute, Academy of Sciences of the Czech Republic, Fri\v{c}ova 298, 25165 Ond\v{r}ejov, Czech Republic}

\author[0000-0002-3963-8701]{Marian Karlick\'{y}}
\affil{Astronomical Institute, Academy of Sciences of the Czech Republic, Fri\v{c}ova 298, 25165 Ond\v{r}ejov, Czech Republic}

\author[0000-0002-7565-5437]{Alena Zemanov\'{a}}
\affil{Astronomical Institute, Academy of Sciences of the Czech Republic, Fri\v{c}ova 298, 25165 Ond\v{r}ejov, Czech Republic}

\author[0000-0002-9690-8456]{Juraj L\"{o}rin\v{c}\'{i}k}
\affil{Bay Area Environmental Research Institute, NASA Research Park, Moffett Field, CA 94035, USA}
\affil{Lockheed Martin Solar \& Astrophysics Laboratory, 3251 Hanover St, Palo Alto, CA 94304, USA}

\author[0000-0001-7312-2410]{Miloslav Druckm\"{u}ller}
\affil{Faculty of Mechanical Engineering, Brno University of Technology, Technick\'{a} 2, 616 69 Brno, Czech Republic}


\begin{abstract}
We report on interaction of the legs of the erupting filament of 2012 August 31 and associated prominent supra-arcade downflows (P-SADs) as observed by the Atmospheric Imaging Assembly onboard the Solar Dynamics Observatory. We employ a number of image processing techniques to enhance weak interacting features. As the filament erupts, both legs stretch outwards. The positive-polarity leg also untwists and splits into two parts. The first part runs into the conjugate (negative-polarity) leg, tearing it apart. The second part then converges into the remnant of the conjugate leg, after which both weaken and finally disappear.
All these episodes of interaction of oppositely-oriented filament legs are followed by appearance of P-SADs, seen in the on-disk projection to be shaped as loop-tops, along with many weaker SADs. All SADs are preceded by hot supra-arcade downflowing loops. 
This observed evolution is consistent with the three-dimensional \textit{rr--rf} (leg--leg) reconnection, where the erupting flux rope reconnects with itself. In our observations, as well as in some models, the reconnection in this geometry is found to be long-lasting. It plays a substantial role in the evolution of the flux rope of the erupting filament and leads to prominent supra-arcade downflows.
\end{abstract}



%
%
\section{Introduction}
\label{Sect:1}

Eruptive solar flares are long recognized to be consequences of magnetic reconnection of oppositely-oriented field lines into new flux rope field lines and flare loops, as described by the Standard solar flare model in two dimensions (2D) \citep{Carmichael64,Sturrock66,Hirayama74,Kopp76}, later extended to 3D \citep{Aulanier12,Aulanier13,Janvier13,Janvier15}; see also the reviews of \citet{LiT21}, \citet{Pontin22}, and \citet{Kazachenko22}. However, the observations of flares and eruptions revealed some problems with this picture. 
For example, \citet{Howard17} found that cores of only two of the filament-associated three-part CMEs at $>$5\,R$_\odot$ contain filaments that can be traced back to the Sun. Majority of the CMEs associated with a filament had cores which did not resemble the erupting filament. In the flare itself, the newly-appearing flare loops move (shrink) far too slowly to constitute reconnection outflows \citep{Warren11,Liu13}. In the peak and gradual phases of the flare, some contracting flare loops create supra-arcade downflows (SADs), now known to be trailing density voids \citep[e.g.,][and references therein]{Asai04,Warren11,Savage11,Savage12a,Savage12b,Innes14,Hanneman14,Reeves17,Chen17,Longcope18,Xue20,LiZ21,Shen22}. Formation of SADs and their predecessors were attributed to inhomogeneous magnetic fields \citep{Warren11} or bursty reconnection \citep{Savage12a,Xue20}. \citet{Shen22} demonstrated that SADs are indeed indirect results of reconnection outflows, and that the preceding shrinking loops slow down in the "interface region" below the termination shock under the current sheet. Nevertheless, the question remains, what reconnection events are the sources of these supra-arcade downflowing loops (SADLs) and SADs in their wake, and which causes account for SADs of various sizes or speeds? And what happens if the filament material and parts of the flux rope get involved in the flare reconnection?

Topologically, the magnetic field lines involved in an eruptive flare belong either to the erupting flux rope (\textit{r}), arcades of loops in the surrounding corona (\textit{a}) or the newly-created flare loops (\textit{f}). \citet{Aulanier19} recognized that three reconnection geometries take place in their simulation, each resulting in a new flux rope field line and a flare loop. In addition to the \textit{aa--rf} reconnection present in the 2D Standard model, the erupting flux rope can also reconnect either with the surrounding corona (\textit{ar--rf} reconnection) or with itself (\textit{rr--rf} reconnection). The former leads to drift of the flux rope footpoints. The latter increases the poloidal flux (twist) of the erupting flux rope \citep{Aulanier19}. \citet{Jiang21} found that the \textit{rr--rf} reconnection can also decrease the toroidal flux of the rope, and that some field lines undergo \textit{rr--rf} reconnection multiple times, potentially making some self-closed in the corona. Reconnection of flux rope field lines, called leg--leg reconnection, was also present in earlier simulations of eruptions 
\citep{Torok05,Gibson06,Karlicky10,Kliem10}.

The \textit{ar--rf} and the associated drift of flux rope footpoints have been observed in many events \citep{Aulanier19,Zemanova19,Lorincik19b,Lorincik21,Dudik19}. Contrary to that, the only direct observational evidence for \textit{rr--rf} reconnection to date was presented by \citet{Dudik19}, even though the process has likely been witnessed in H$\alpha$ filament eruptions before \citep{Kotrc98}. Here, we report on protracted occurrence of \textit{rr--rf} reconnection in the well-known filament eruption of 2012 August 31. Observations and image processing techniques are summarized in Section \ref{Sect:2}, while the filament eruption and the associated flare are described in Sections \ref{Sect:3} and \ref{Sect:4}, respectively. Finally, interpretation is given in Section \ref{Sect:5}.

\begin{figure*}[!h]
	\centering 

	\includegraphics[width=17.00cm,viewport= 0  0 995 258,clip]{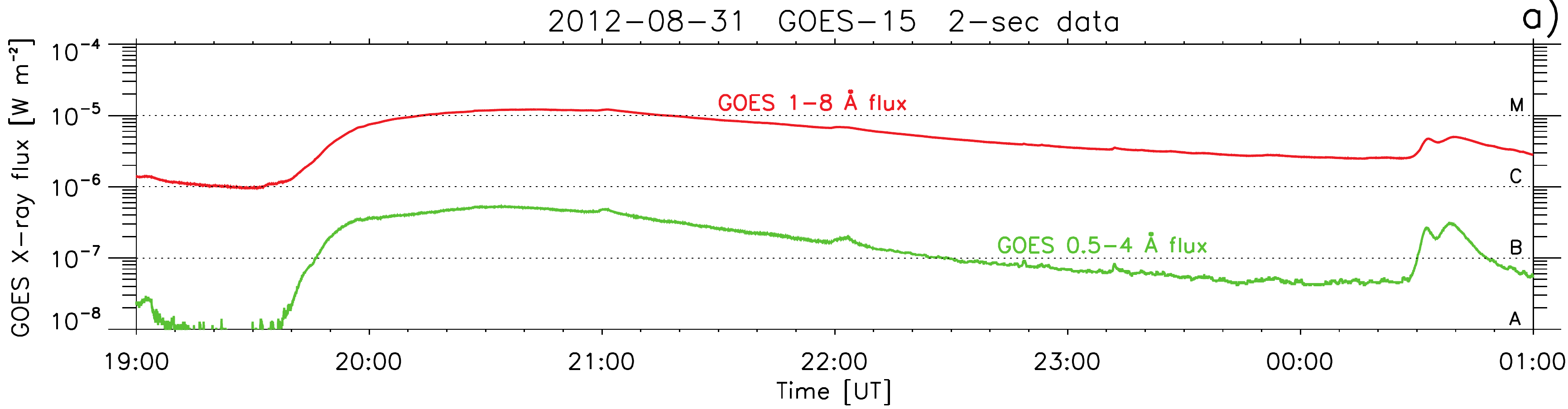}
	\includegraphics[width=9.14cm,viewport= 0 45 495 270,clip]{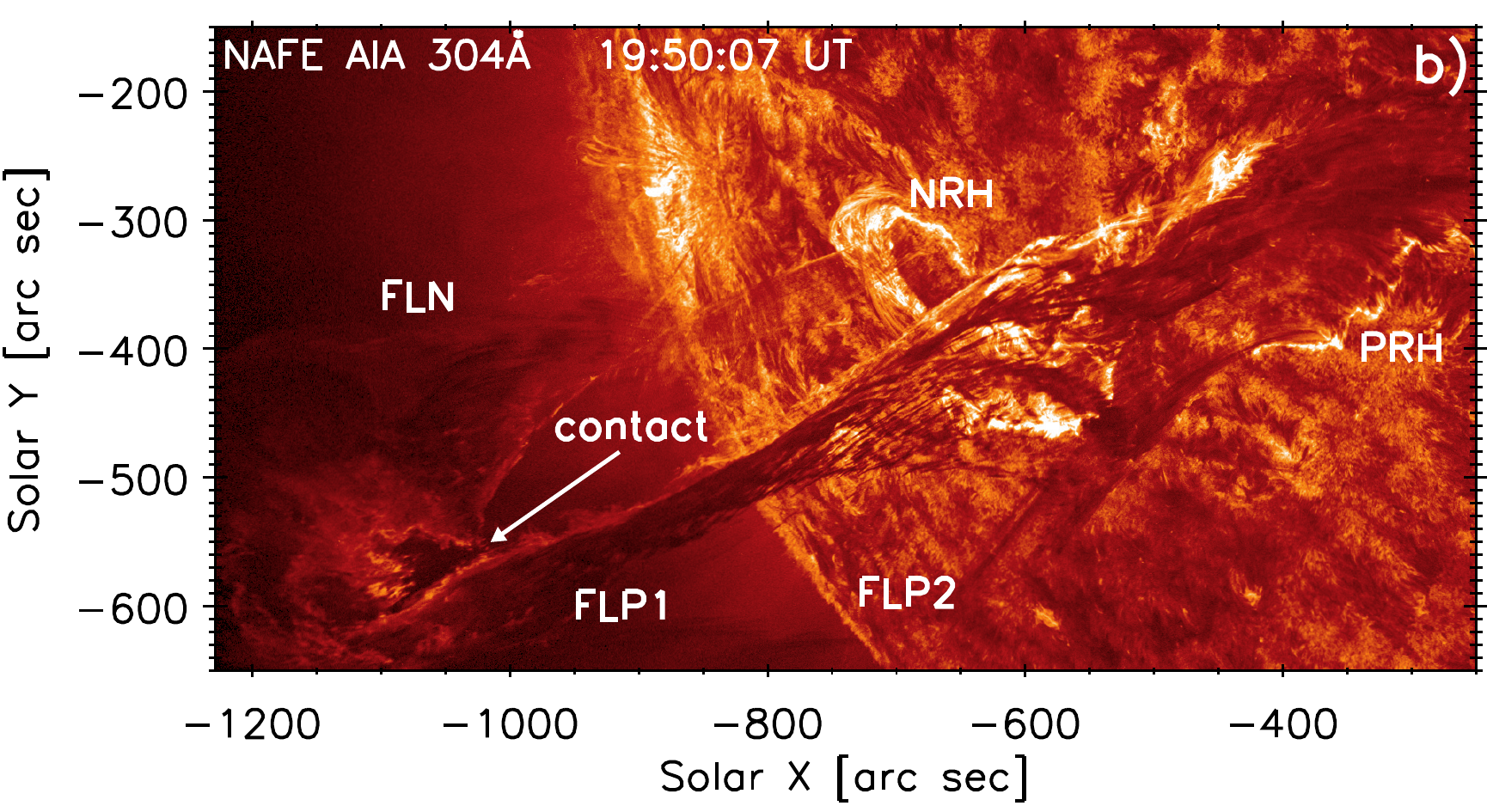}
	\includegraphics[width=7.85cm,viewport=70 45 495 270,clip]{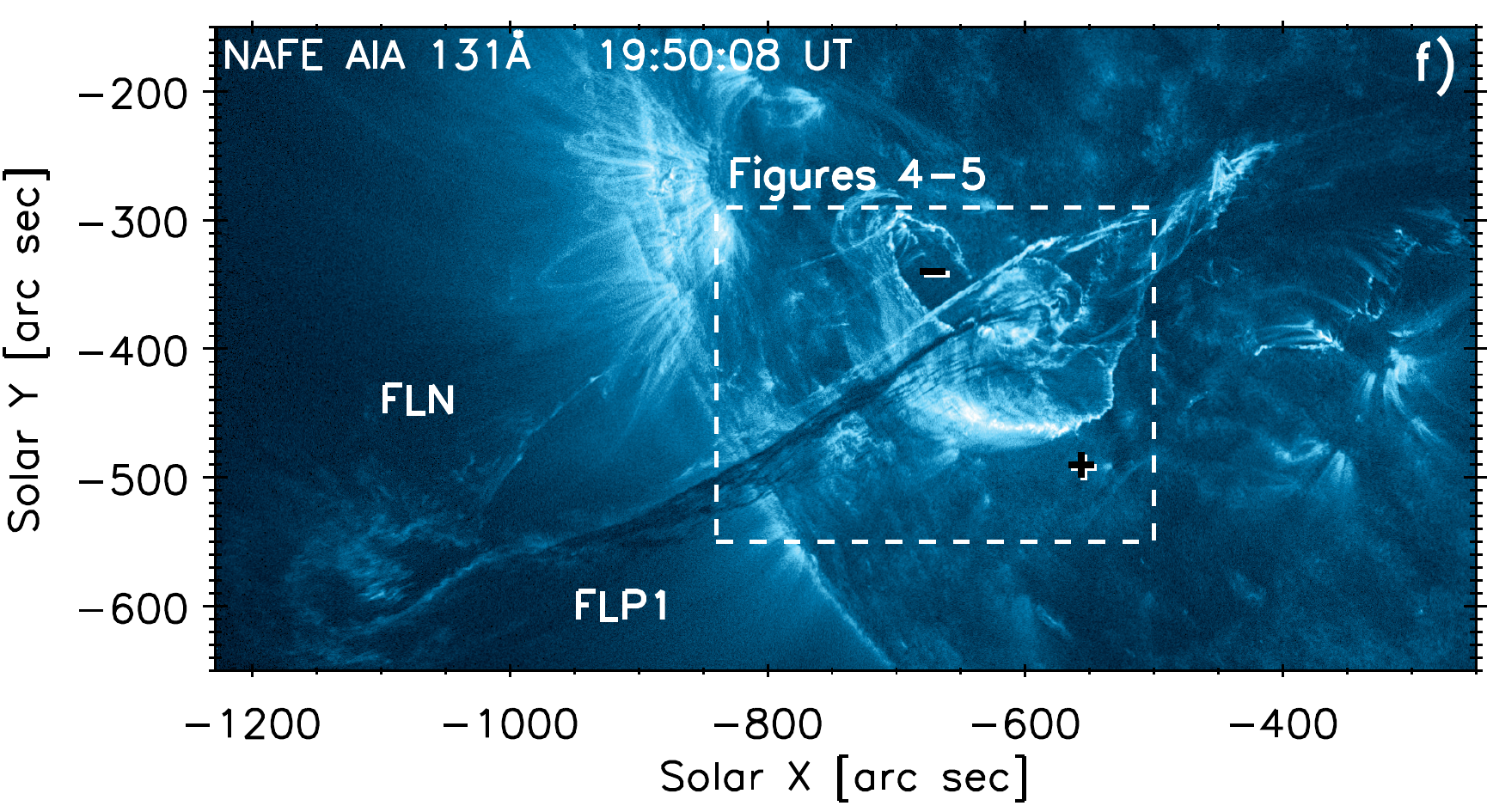}
	\includegraphics[width=9.14cm,viewport= 0 45 495 265,clip]{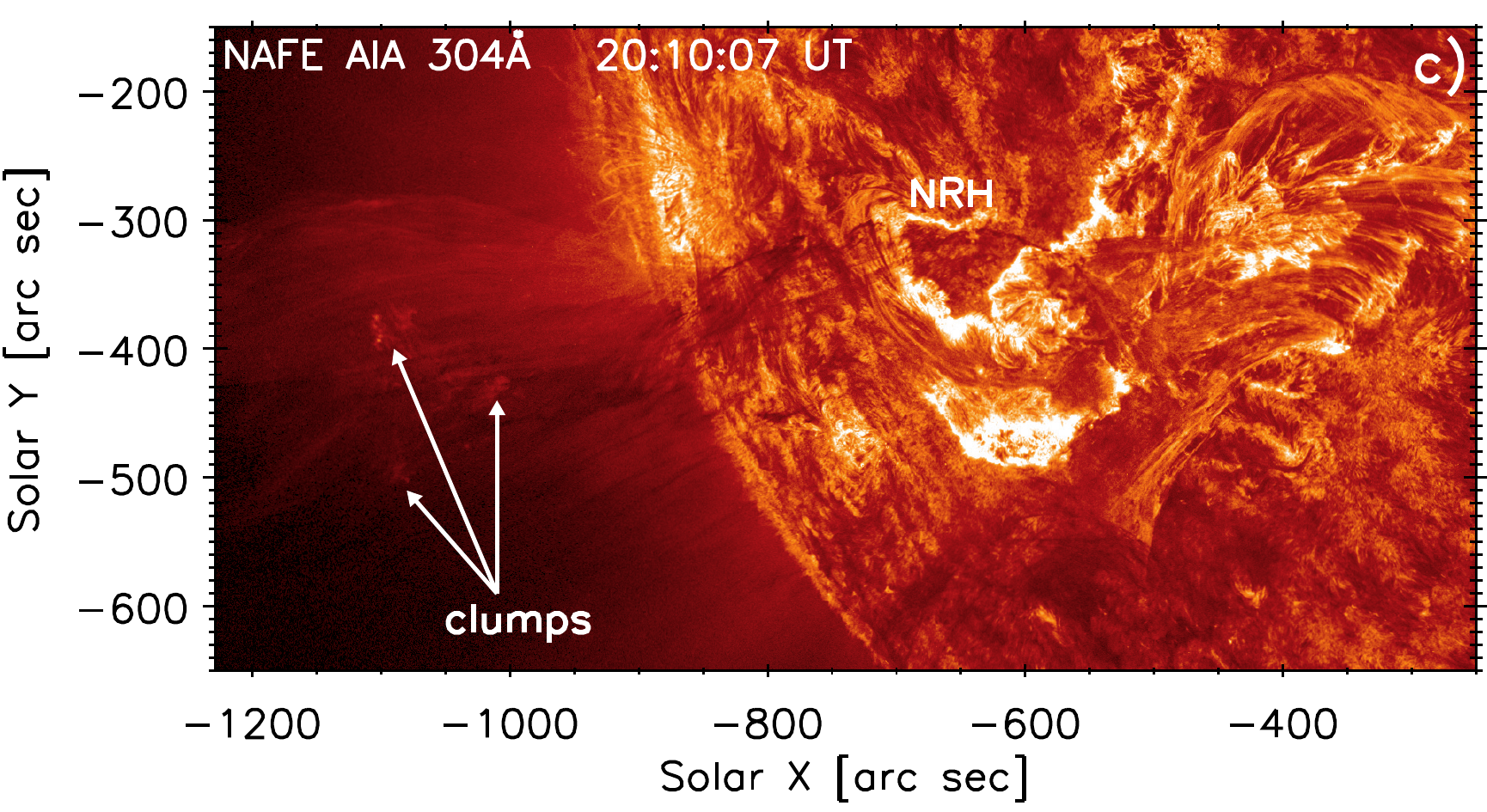}
	\includegraphics[width=7.85cm,viewport=70 45 495 265,clip]{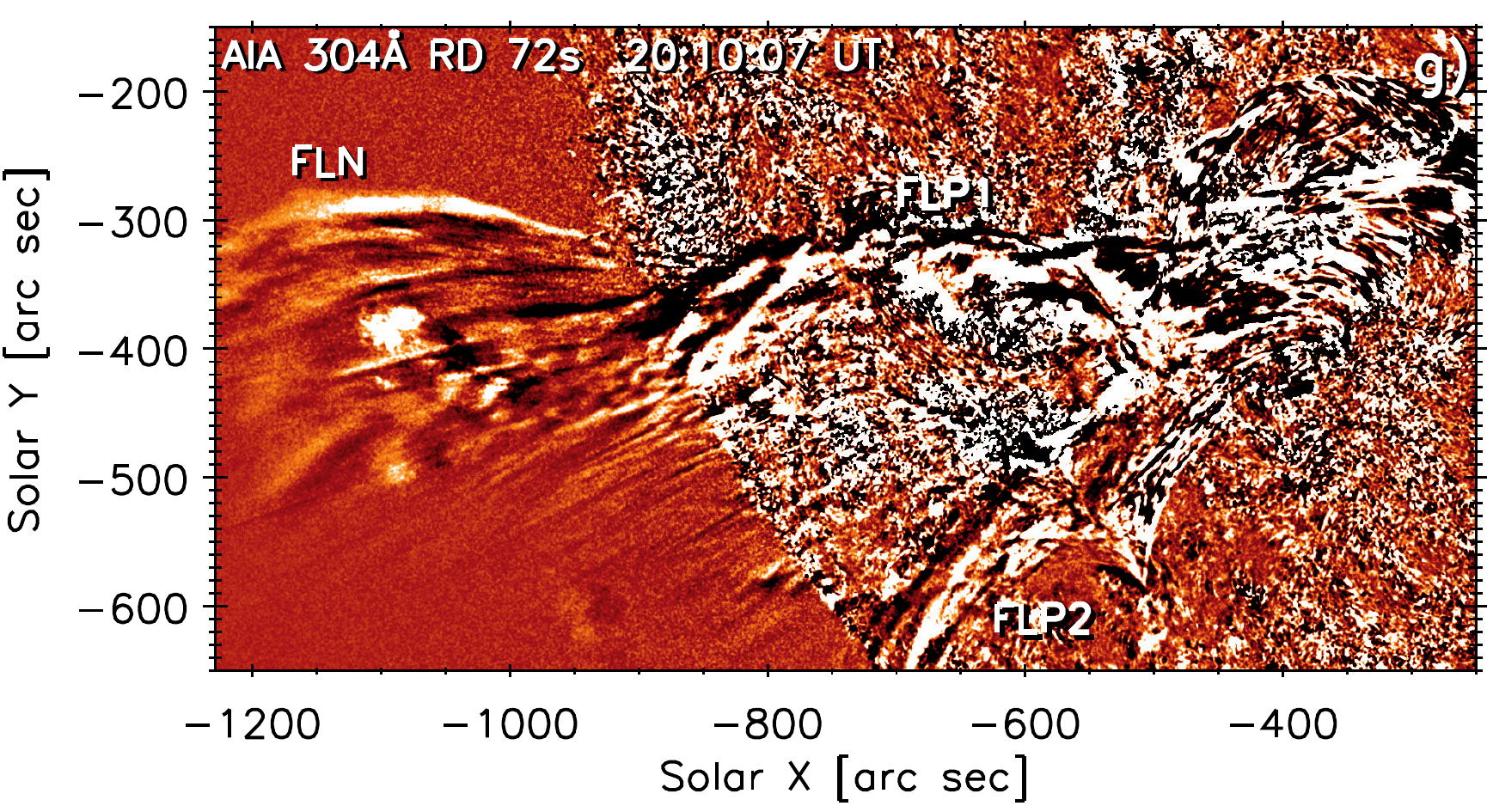}
	\includegraphics[width=9.14cm,viewport= 0 45 495 265,clip]{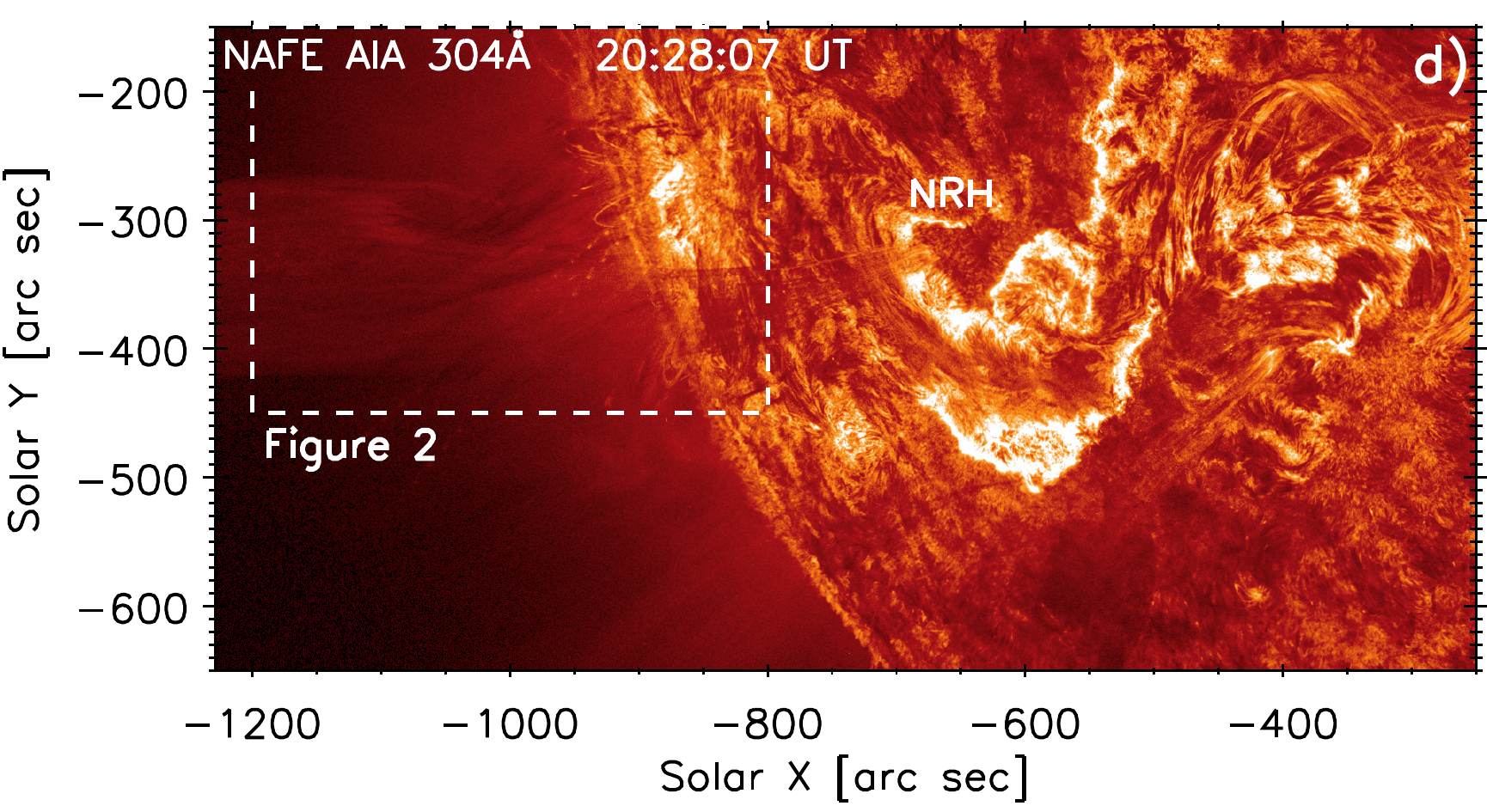}
	\includegraphics[width=7.85cm,viewport=70 45 495 265,clip]{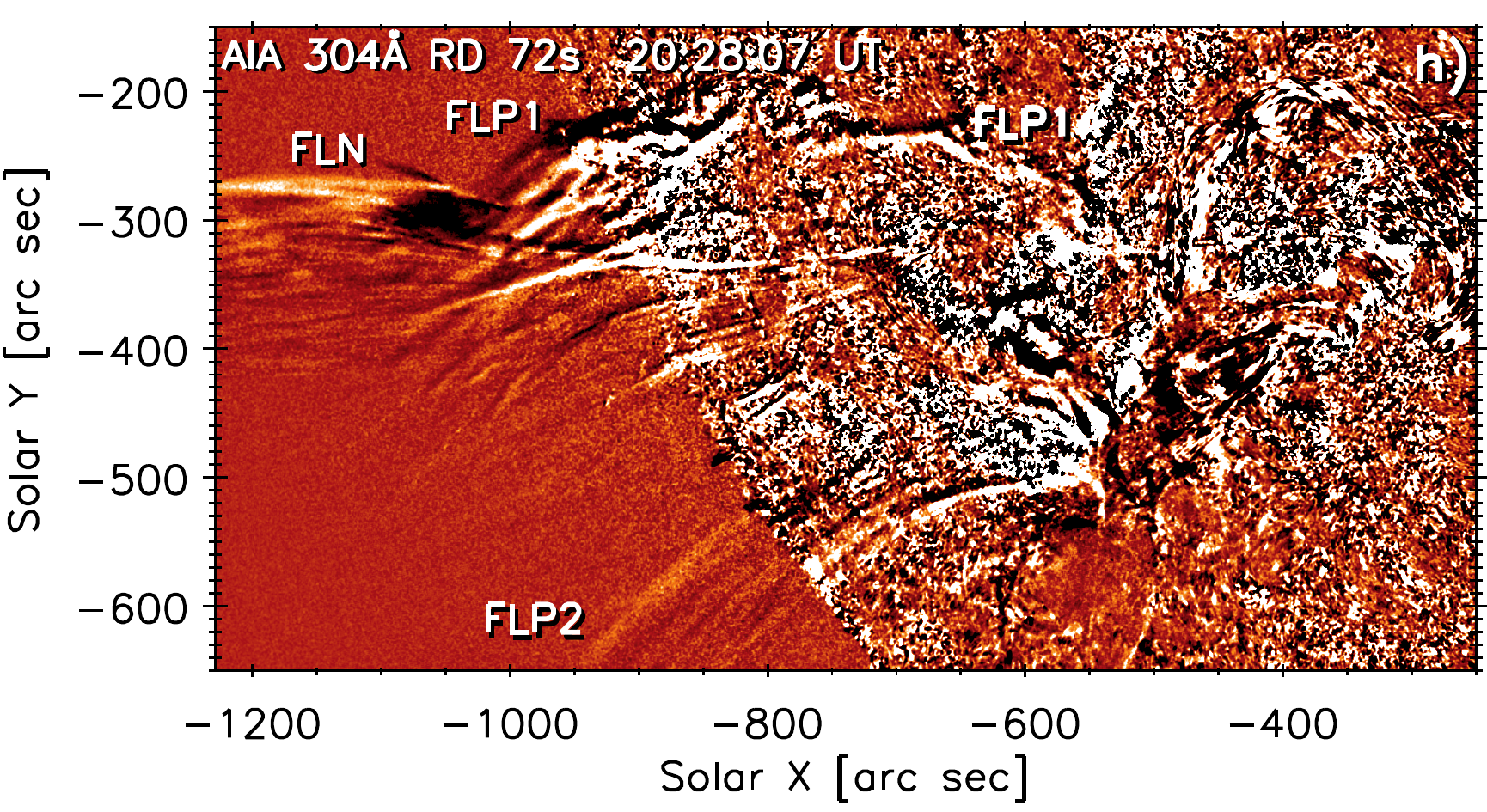}
	\includegraphics[width=9.14cm,viewport= 0  0 495 265,clip]{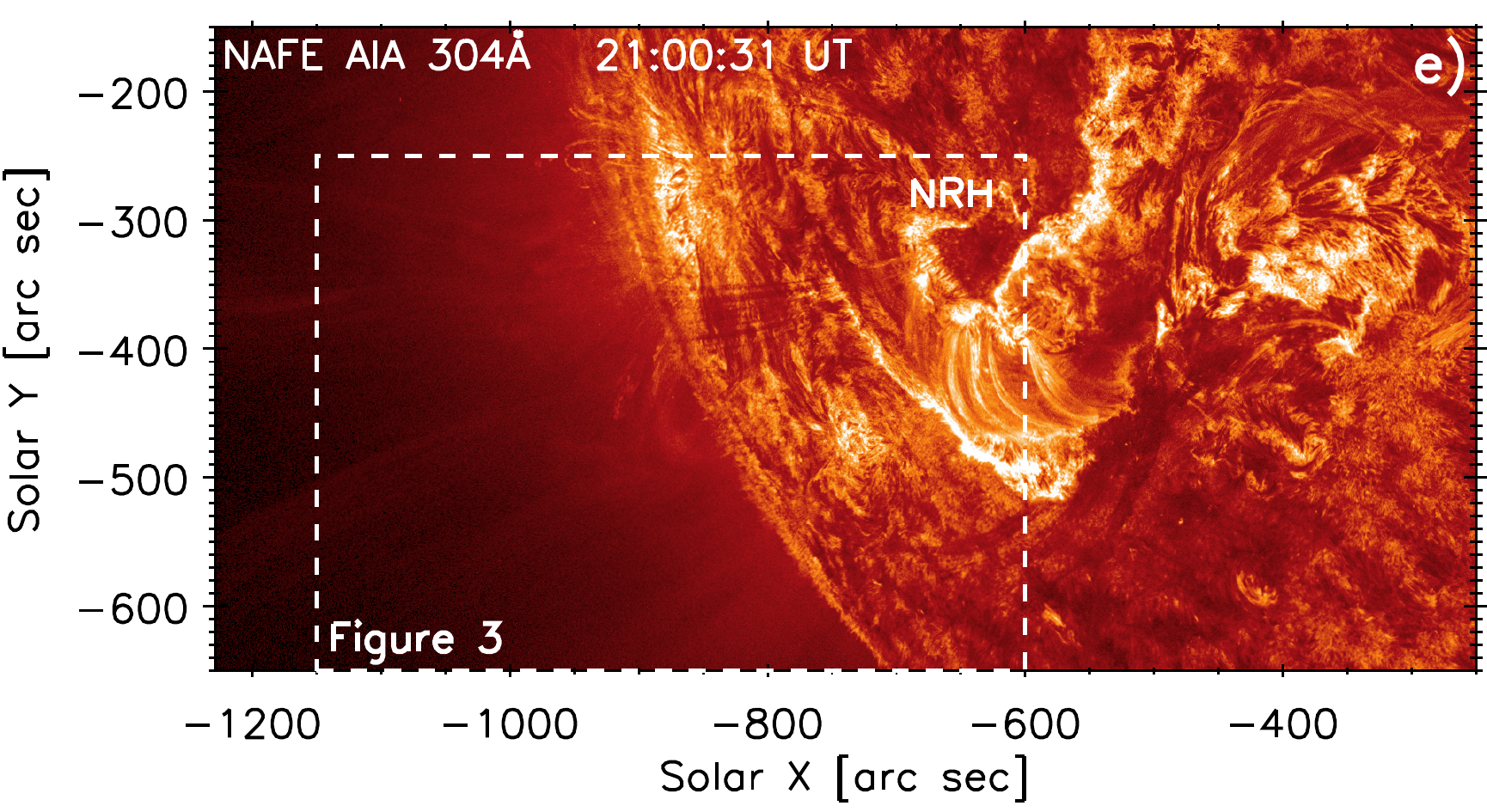}
	\includegraphics[width=7.85cm,viewport=70  0 495 265,clip]{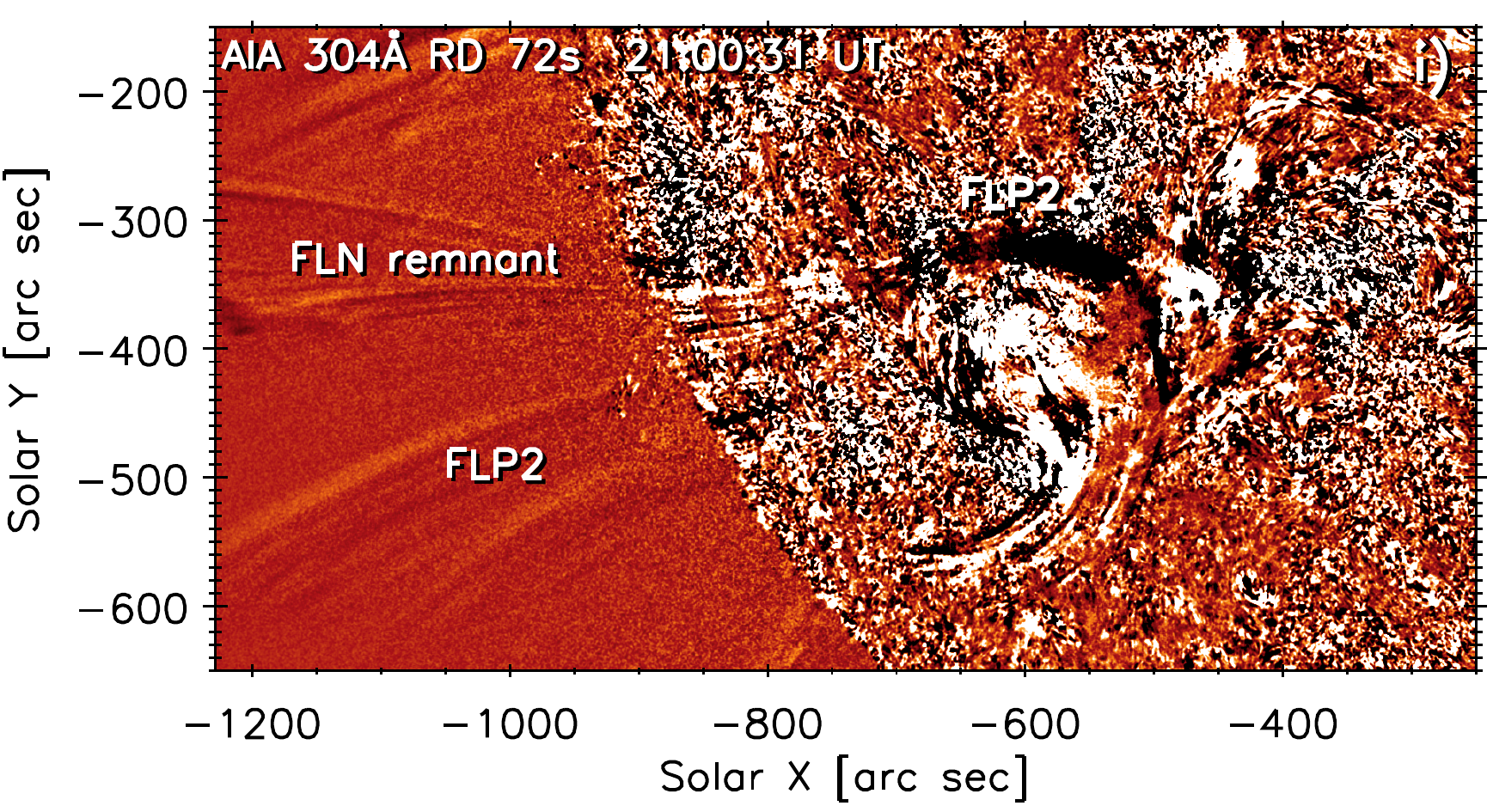}
\caption{The 2012 August 31 event. (a): \textit{GOES} X-ray flux. (b--e): Filament eruption as observed in the NAFE-processed AIA 304\,\AA. (f): Flare loops below the filament seen in 131\,\AA. Dominant polarities on both sides of flare are indicated. (g--i): Running-difference 304\,\AA~observations. Evolving features discussed in the text are indicated. An animation of panels b--e, f, and g--i is available, spanning 19:30--22:00\,UT. Its real-time duration is 42\,s.}
\label{Fig:Eruption}
\end{figure*}
%
%
\begin{figure*}
	\centering
 	\includegraphics[width=5.41cm,viewport= 0 43 280 205,clip]{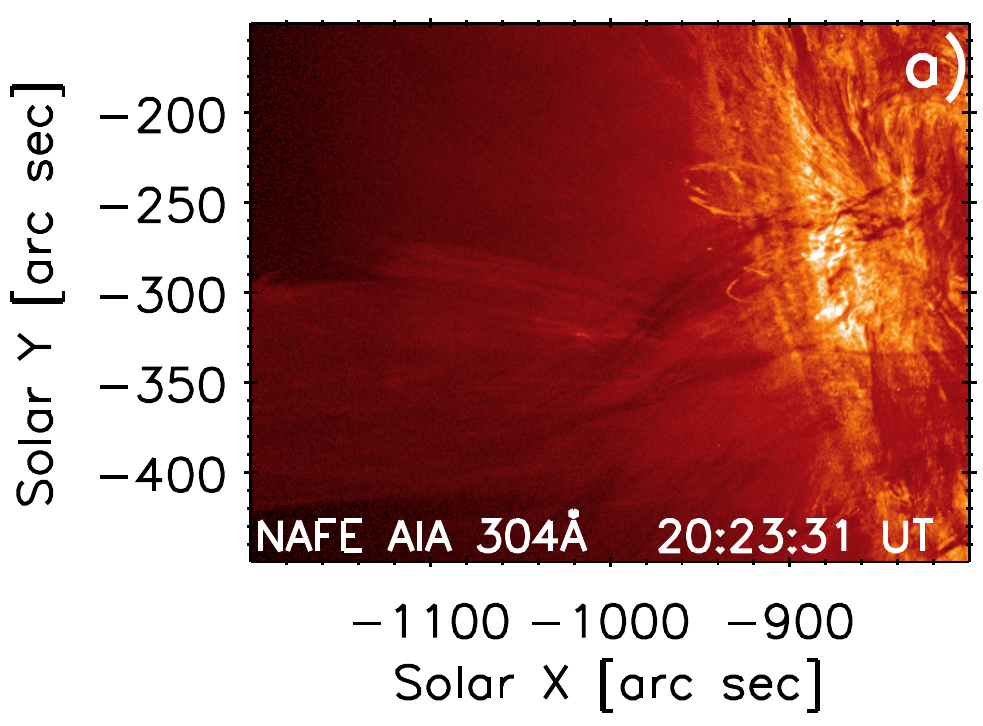}
	\includegraphics[width=4.06cm,viewport=70 43 280 205,clip]{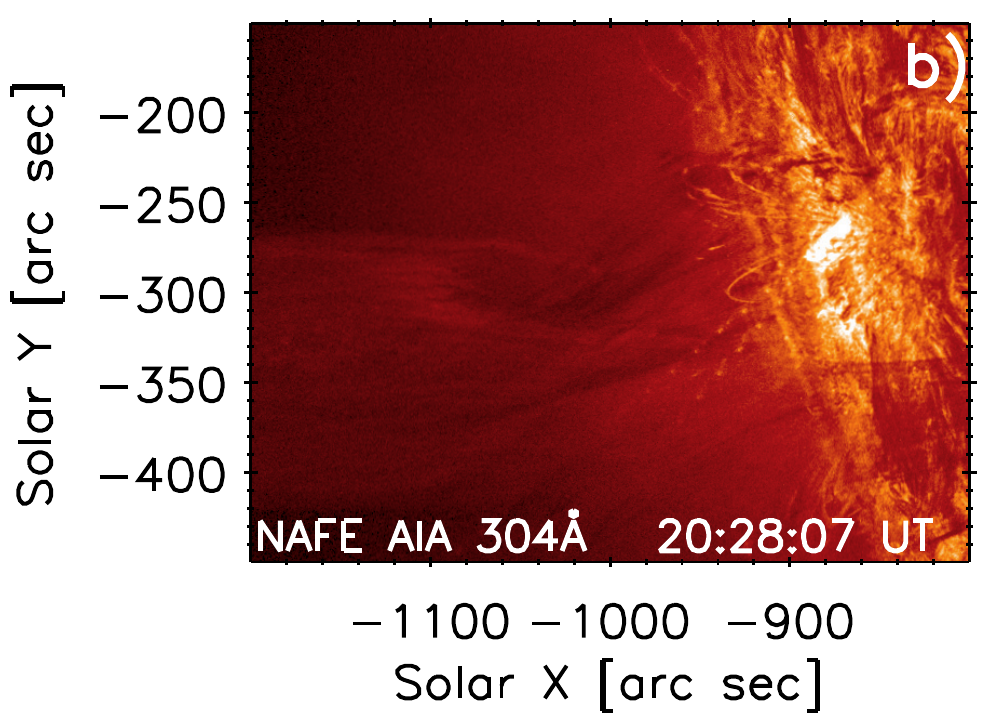}
	\includegraphics[width=4.06cm,viewport=70 43 280 205,clip]{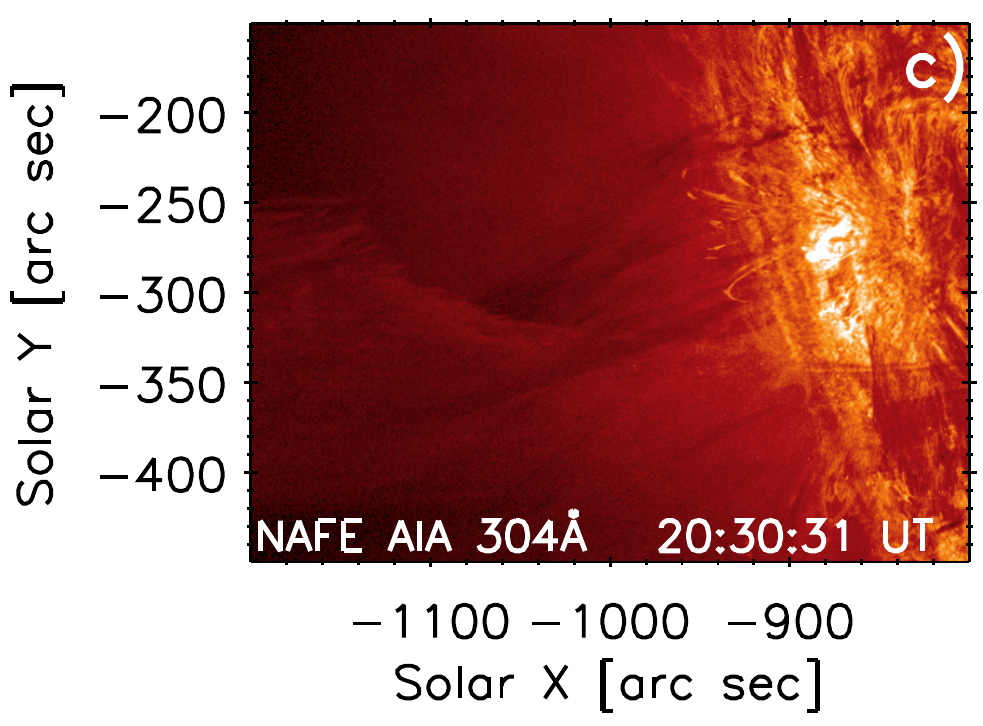}
	\includegraphics[width=4.06cm,viewport=70 43 280 205,clip]{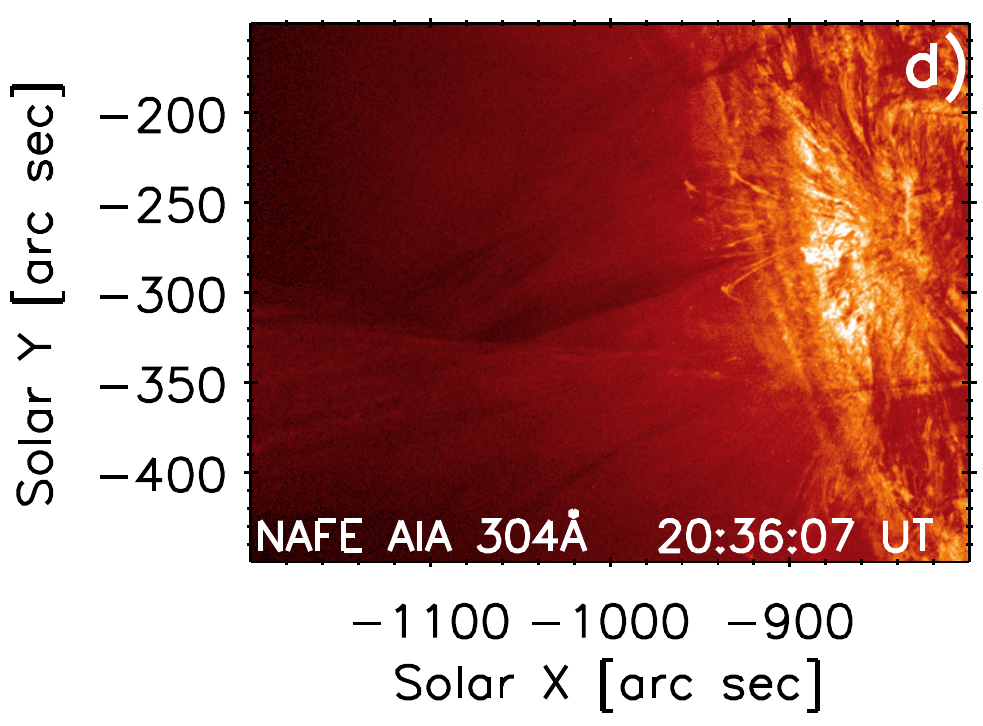}
 	\includegraphics[width=5.41cm,viewport= 0  0 280 205,clip]{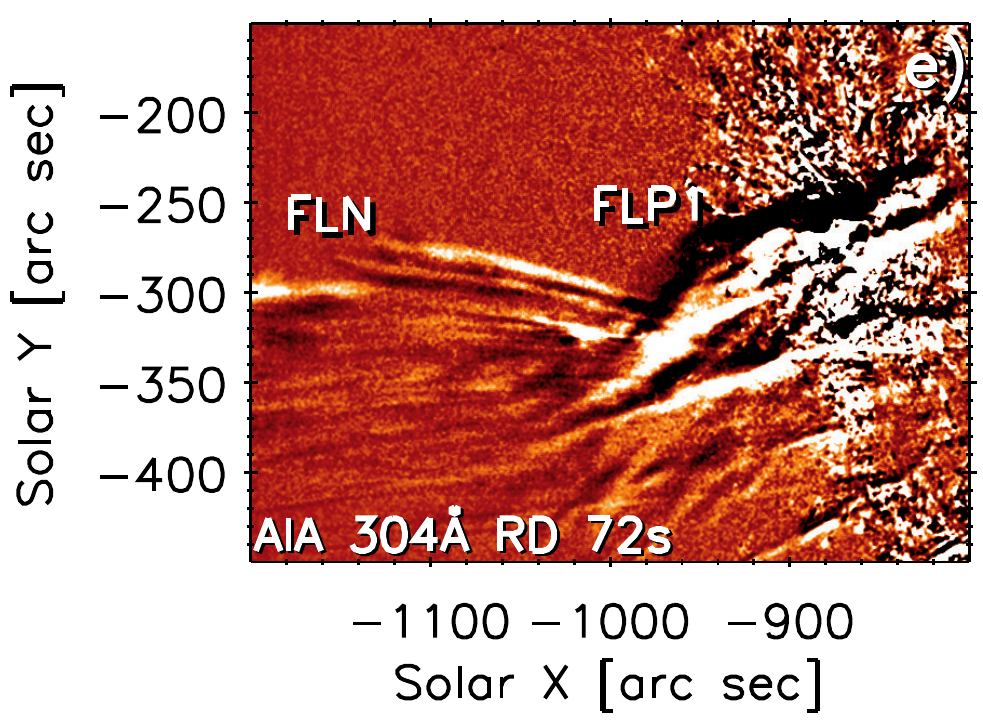}
	\includegraphics[width=4.06cm,viewport=70  0 280 205,clip]{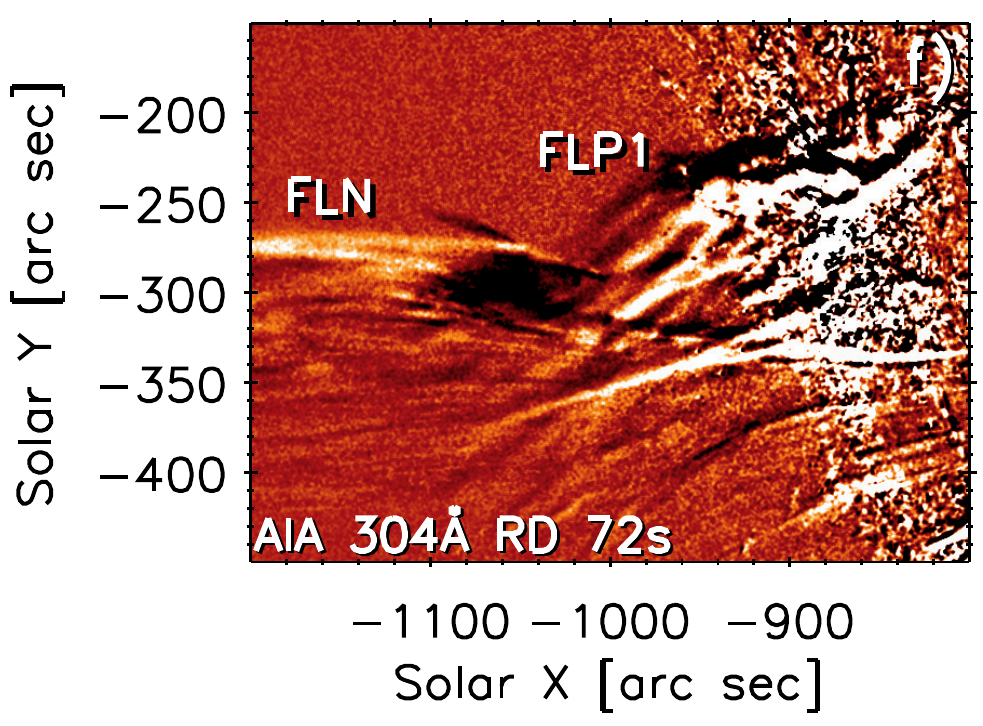}
	\includegraphics[width=4.06cm,viewport=70  0 280 205,clip]{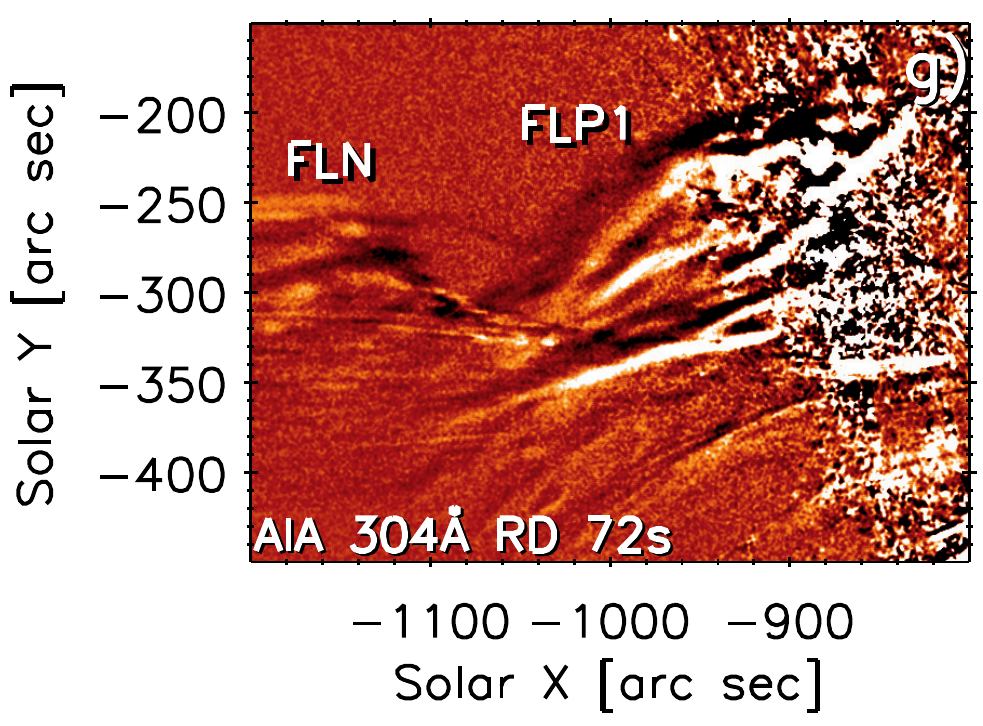}
	\includegraphics[width=4.06cm,viewport=70  0 280 205,clip]{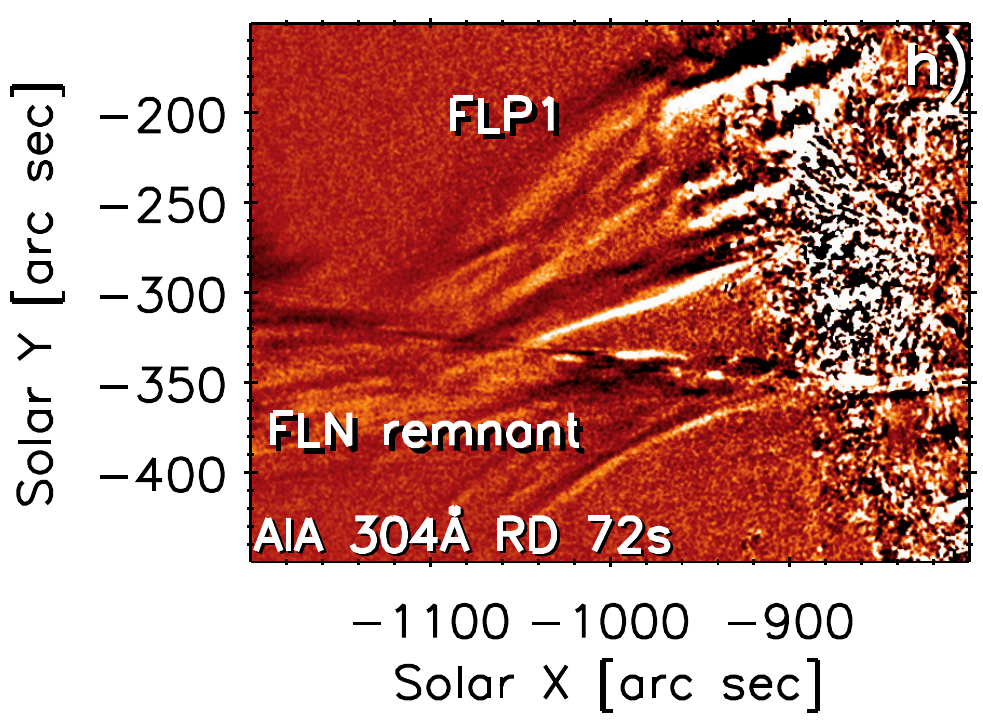}
\caption{Evolution of the filament at around 20:28 UT. Top row (a--d) shows the NAFE 304\,\AA~images, while the bottom row (panels e--h) show the 72 second running difference (RD) 304\,\AA~images. 
An animation is available, spanning 20:20--20:40\,UT. Its real-time duration is 5\,s.}
\label{Fig:aia304_2028UT}
\end{figure*}
%
\begin{figure*}
	\centering
 	\includegraphics[width=5.41cm,viewport= 0 43 280 205,clip]{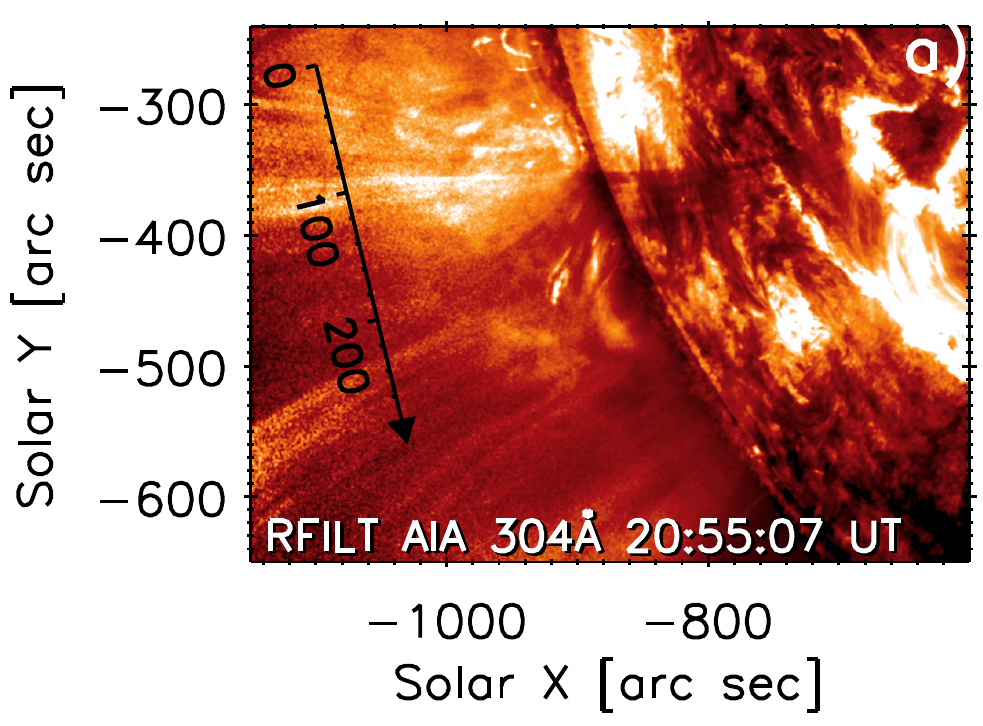}
	\includegraphics[width=4.06cm,viewport=70 43 280 205,clip]{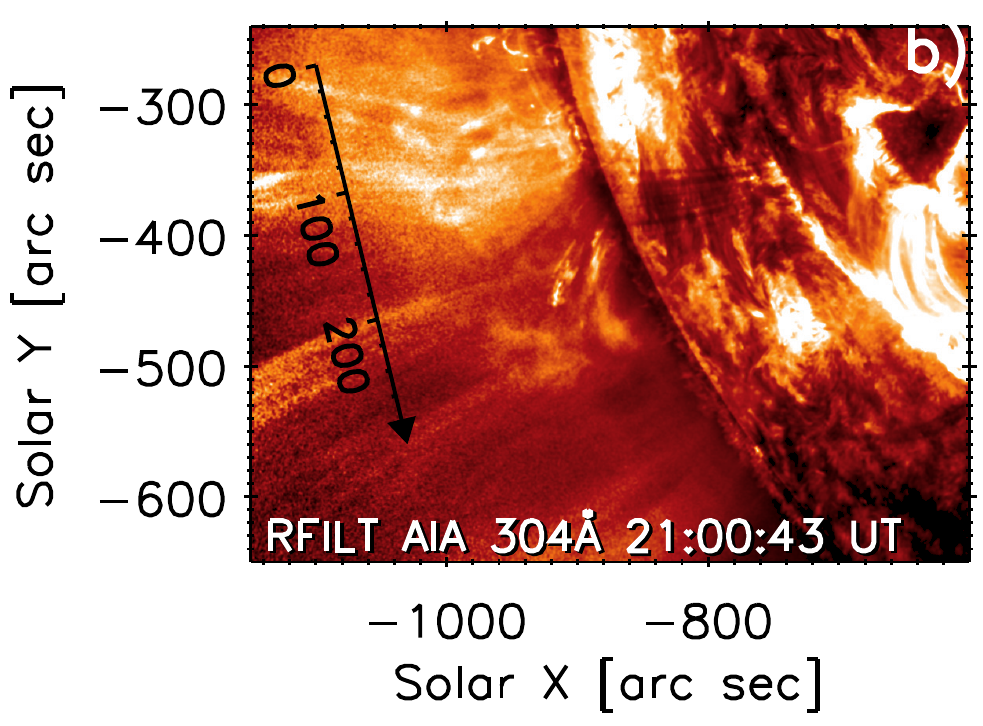}
	\includegraphics[width=4.06cm,viewport=70 43 280 205,clip]{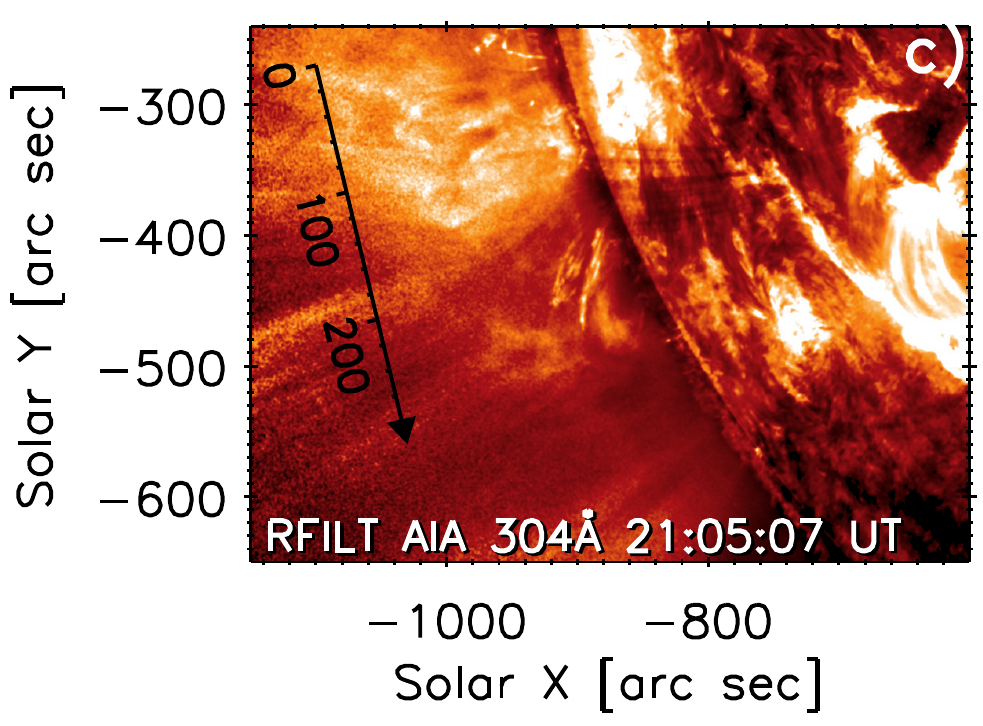}
	\includegraphics[width=4.06cm,viewport=70 43 280 205,clip]{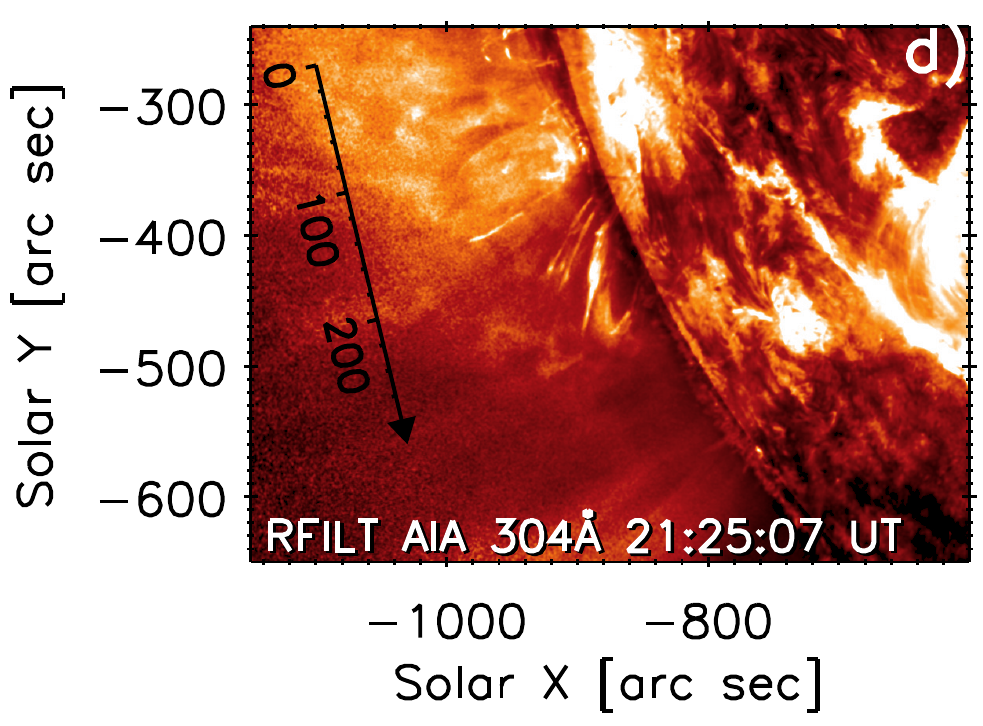}
 	\includegraphics[width=5.41cm,viewport= 0  0 280 205,clip]{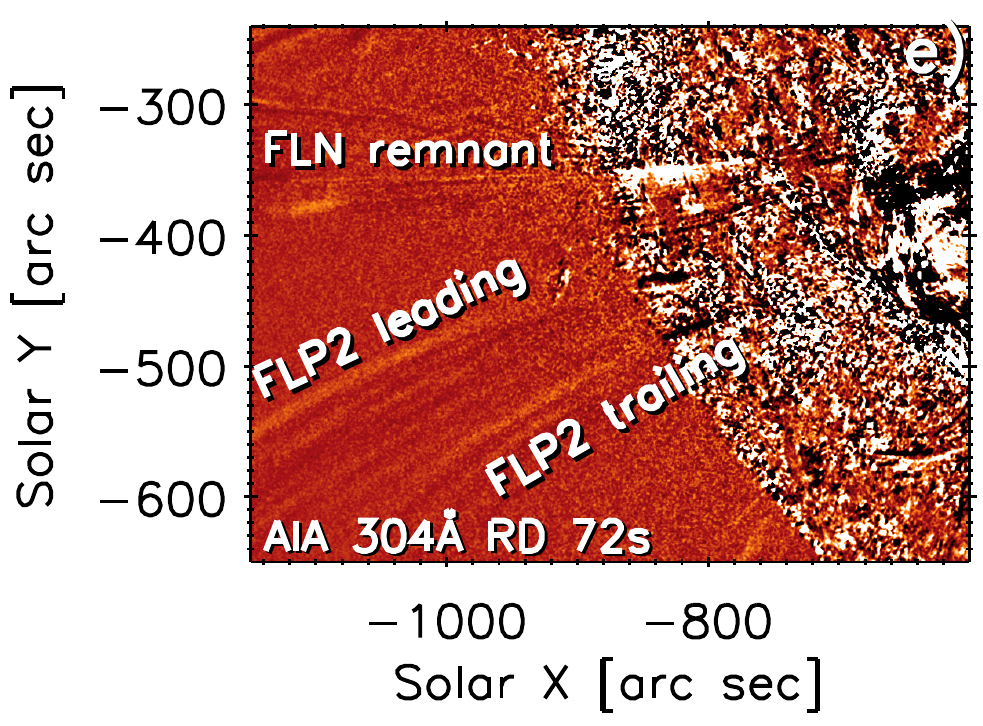}
	\includegraphics[width=4.06cm,viewport=70  0 280 205,clip]{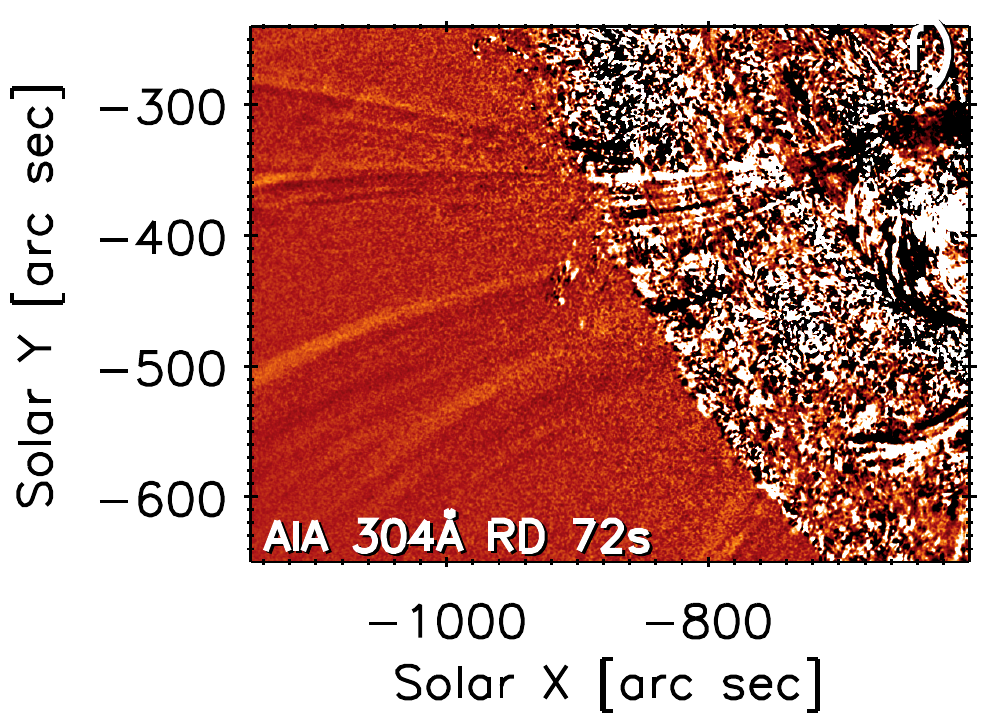}
	\includegraphics[width=4.06cm,viewport=70  0 280 205,clip]{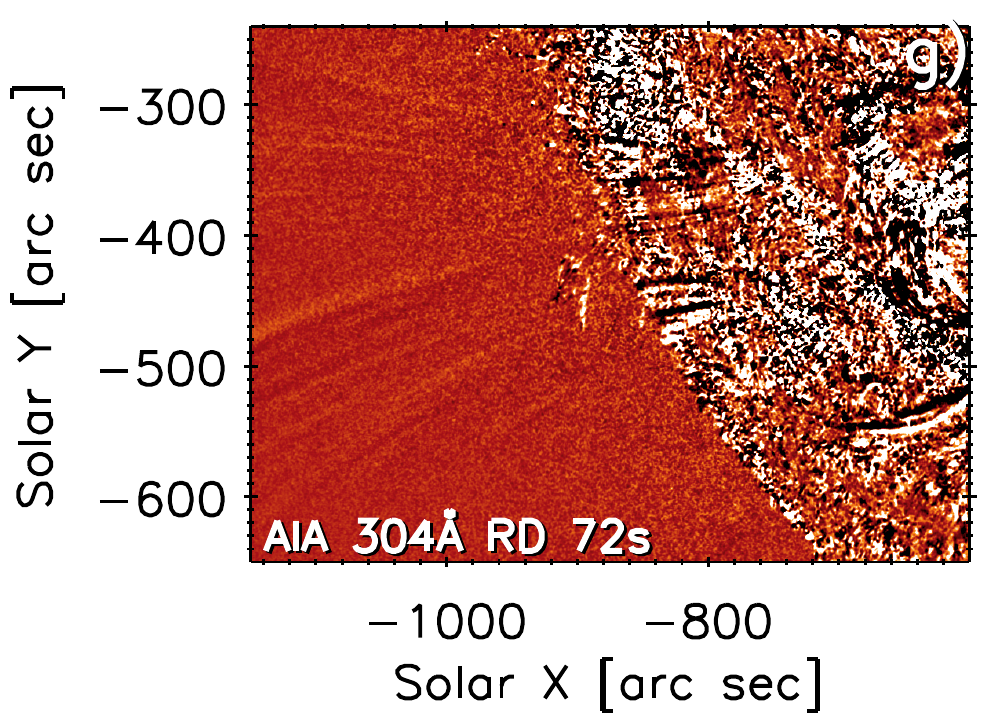}
	\includegraphics[width=4.06cm,viewport=70  0 280 205,clip]{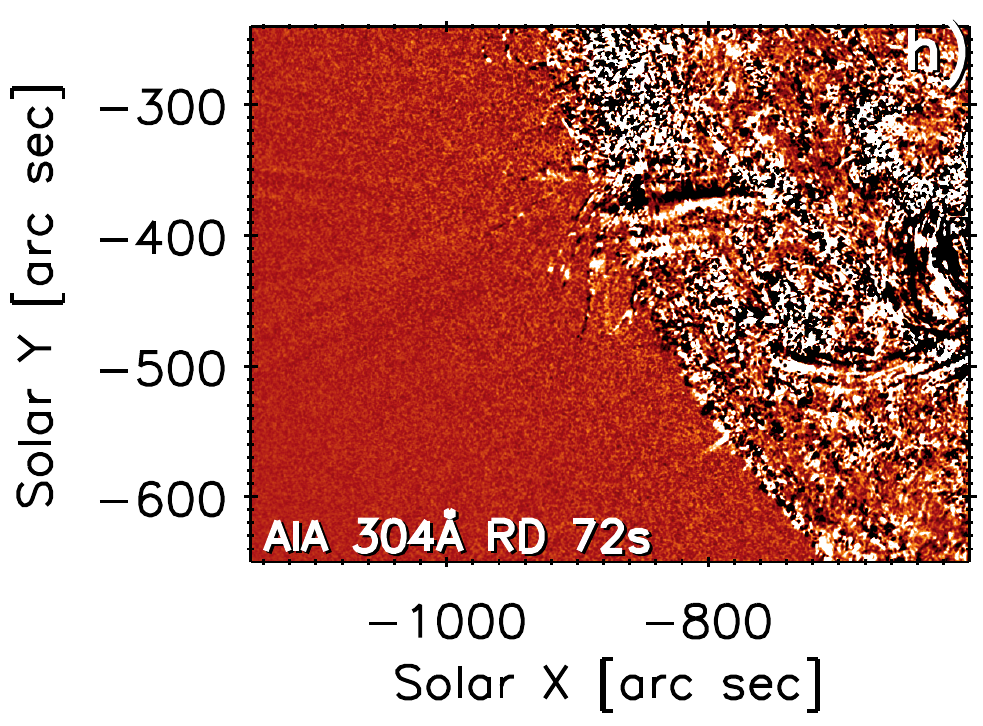}
	\includegraphics[width=17.6cm,clip]{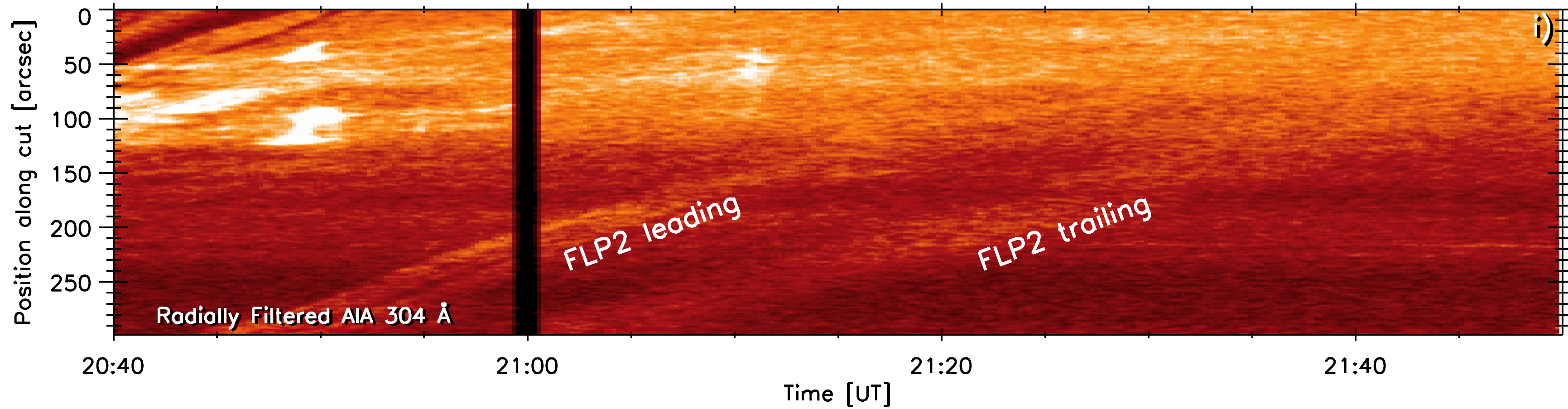}
\caption{Evolution of the filament at around 21:00 UT. Top row (a--d) shows radially filtered AIA 304\,\AA~images, while the bottom row (panels e--h) show the 72 second running difference (RD) 304\,\AA~images. 
Panel (i) shows the time-distance plot along the cut shown in panels (a)--(d). An animation is available, spanning 20:50--22:00\,UT. Its real-time duration is 14\,s.}
\label{Fig:aia304_2100UT}
\end{figure*}
%
%
%
\begin{figure*}
	\centering
 	\includegraphics[width=5.41cm,viewport= 0  0 280 205,clip]{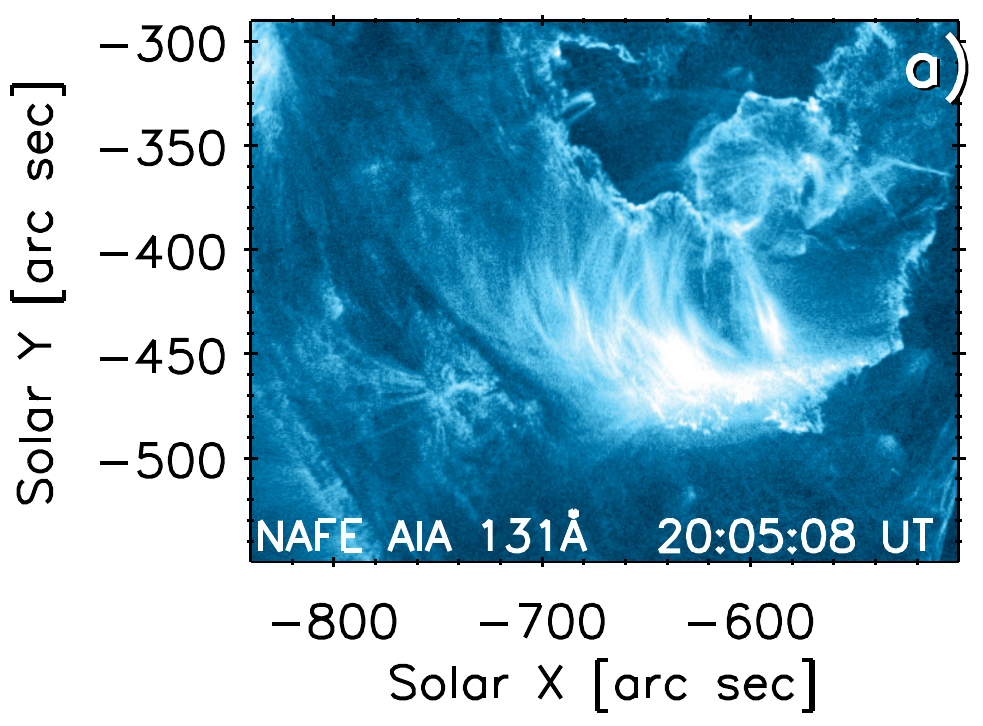}
	\includegraphics[width=4.06cm,viewport=70  0 280 205,clip]{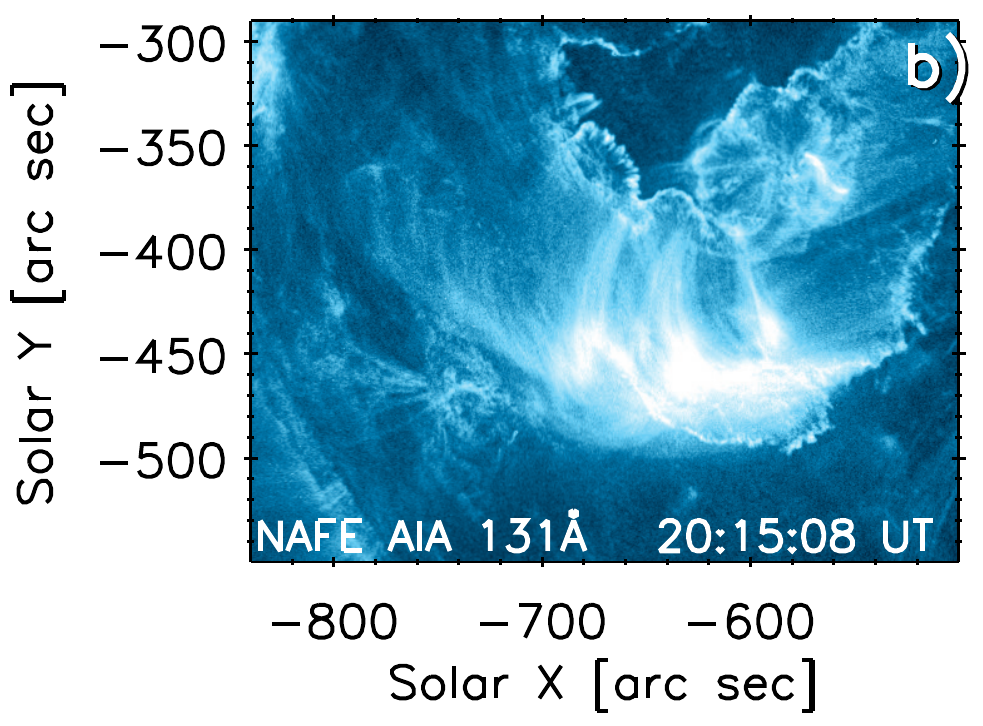}
	\includegraphics[width=4.06cm,viewport=70  0 280 205,clip]{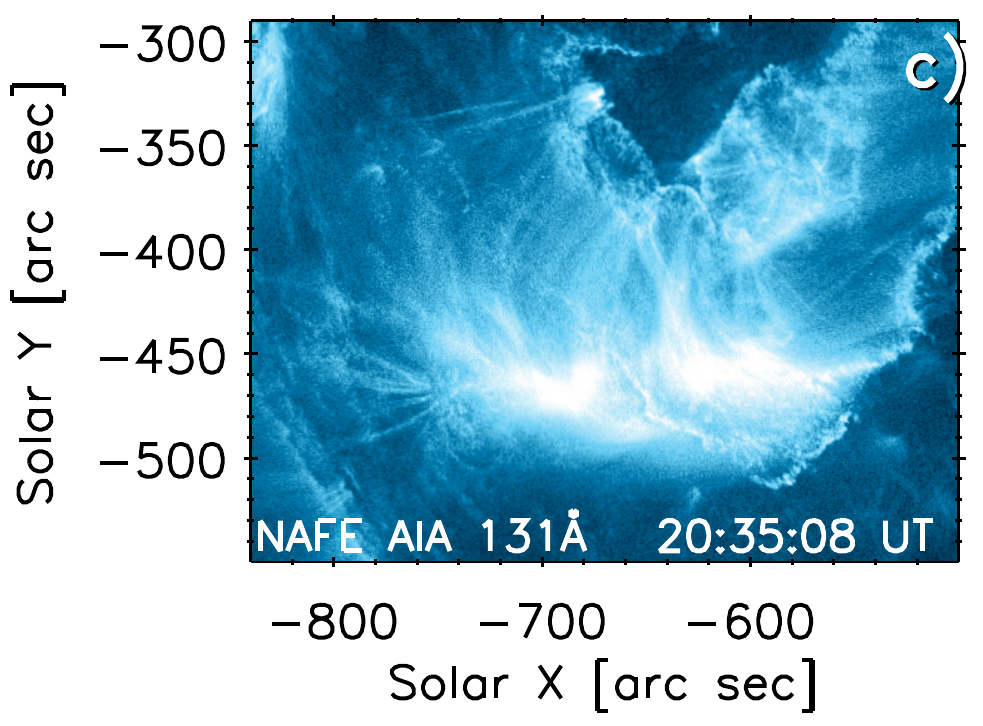}
	\includegraphics[width=4.06cm,viewport=70  0 280 205,clip]{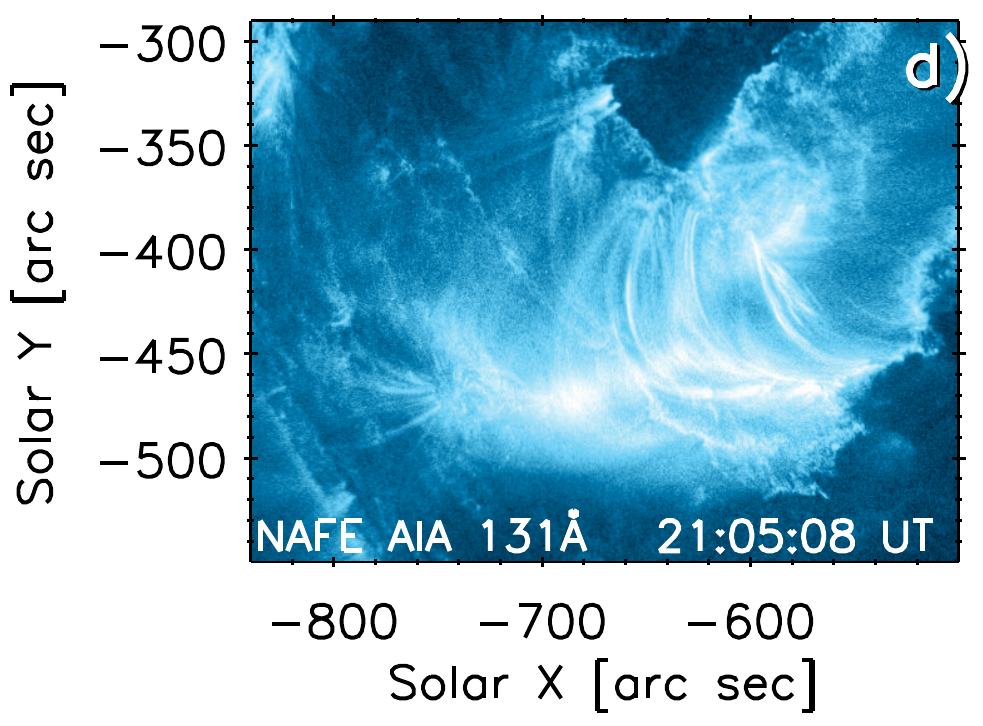}
	\includegraphics[width= 5.50cm,viewport= 0 43 285 205,clip]{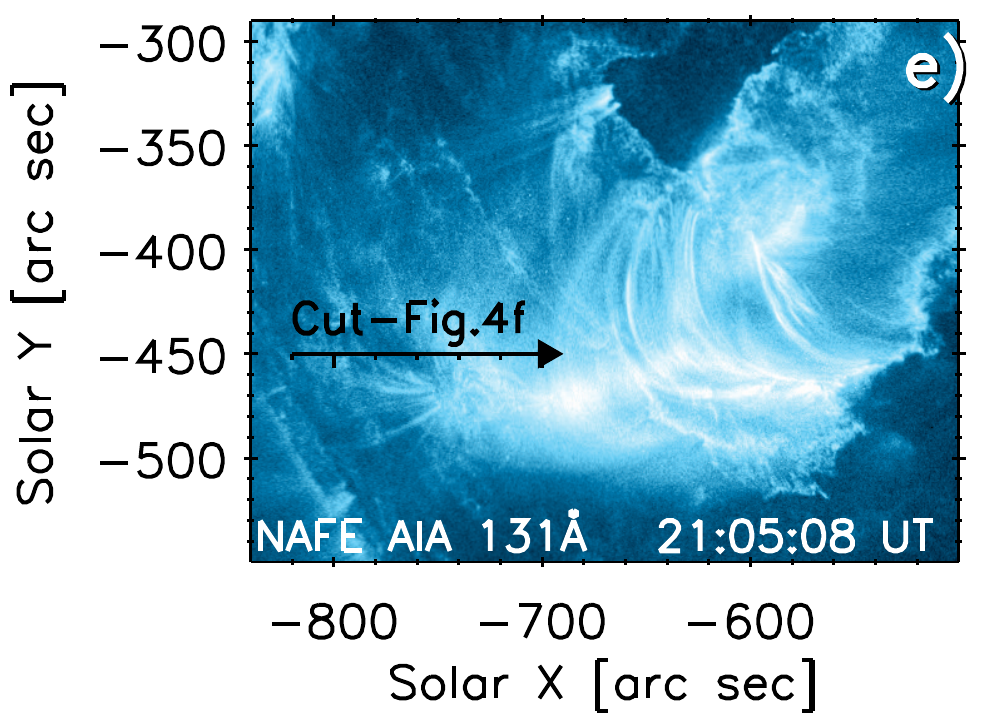}
	\includegraphics[width=12.09cm,viewport= 0 43 566 200,clip]{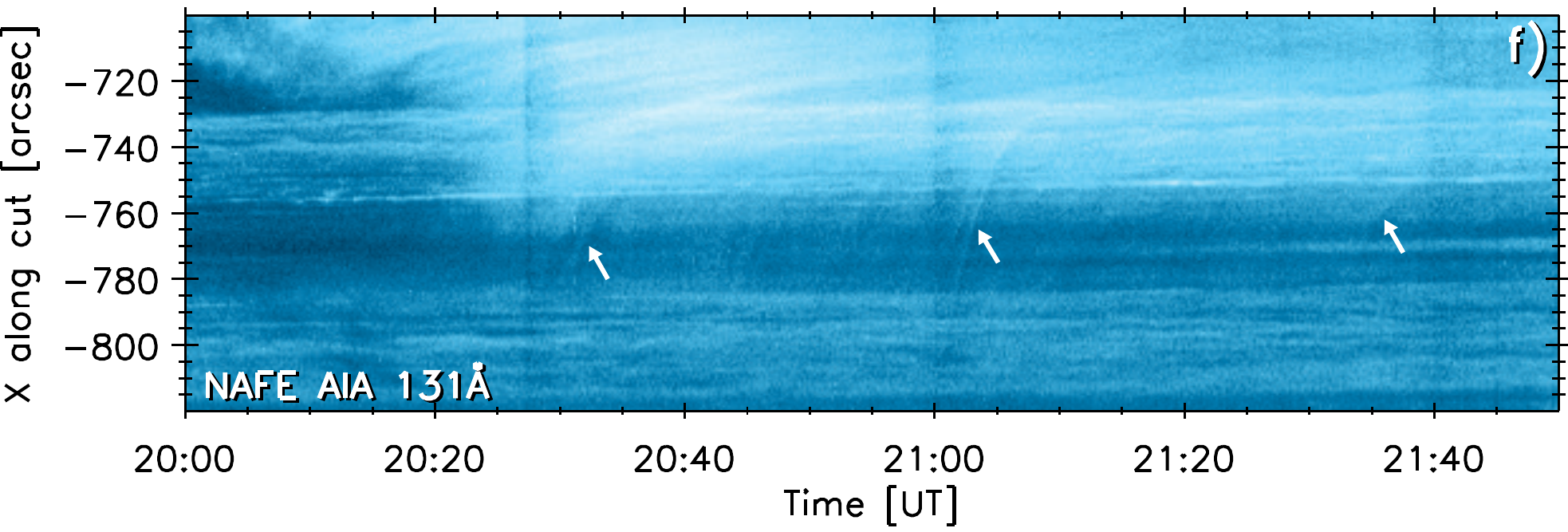}
	\includegraphics[width= 5.50cm,viewport= 0  3 285 205,clip]{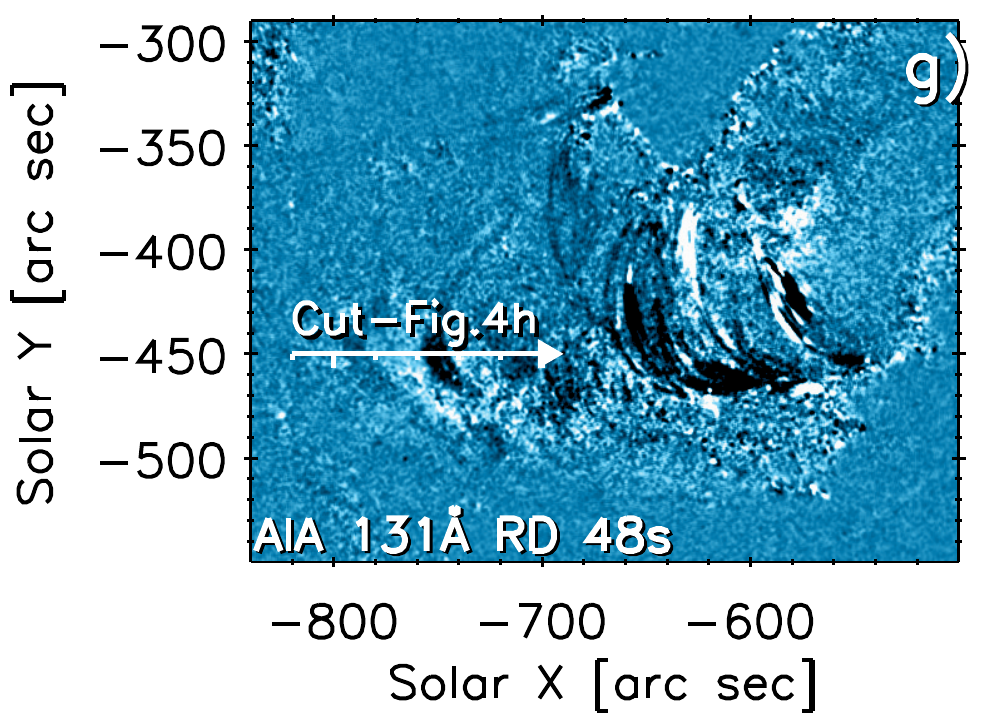}
	\includegraphics[width=12.09cm,viewport= 0  3 565 200,clip]{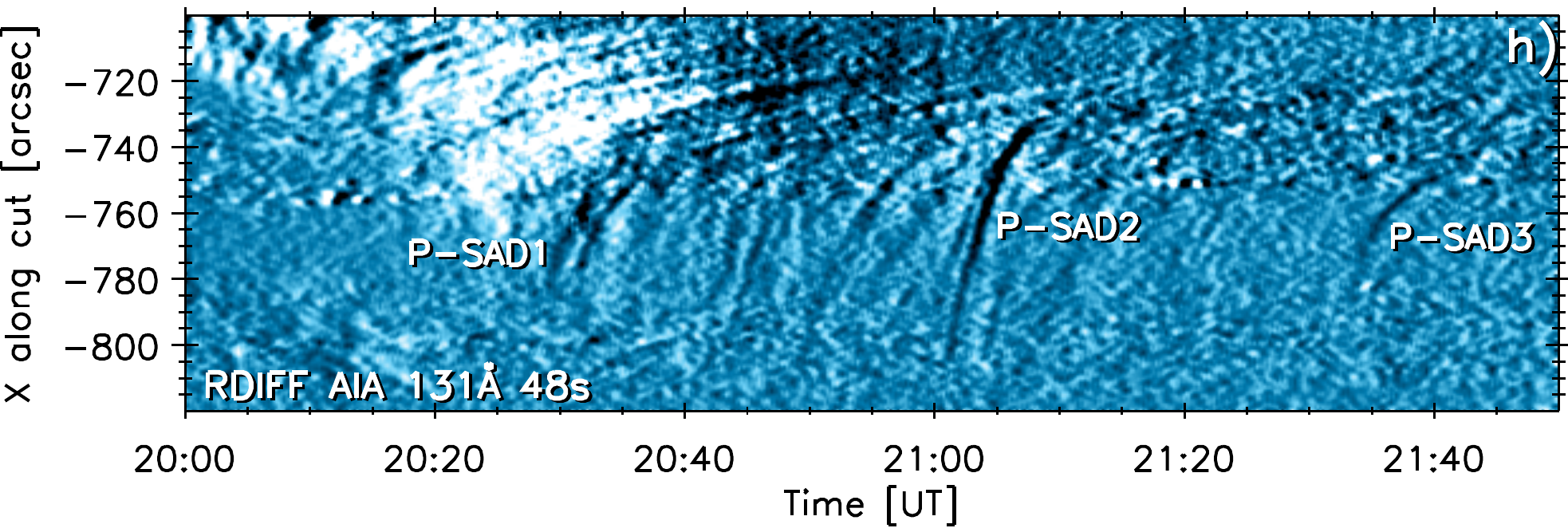}
\caption{Evolution of the supra-arcade region (a--d). Time-distance diagrams constructed along the cut shown in panels (e) and (g) are shown for both NAFE and RD data in panels (f) and (h), respectively. White arrows in the panel f point to supra-arcade downflowing loops preceding P-SADs1--3. An animation is available, spanning the time interval 19:50--22:00\,UT. Its real-time duration is 32\,s.}
\label{Fig:aia131_arcade}
\end{figure*}
%
\begin{figure*}
	\centering
 	\includegraphics[width=5.41cm,viewport= 0  0 280 205,clip]{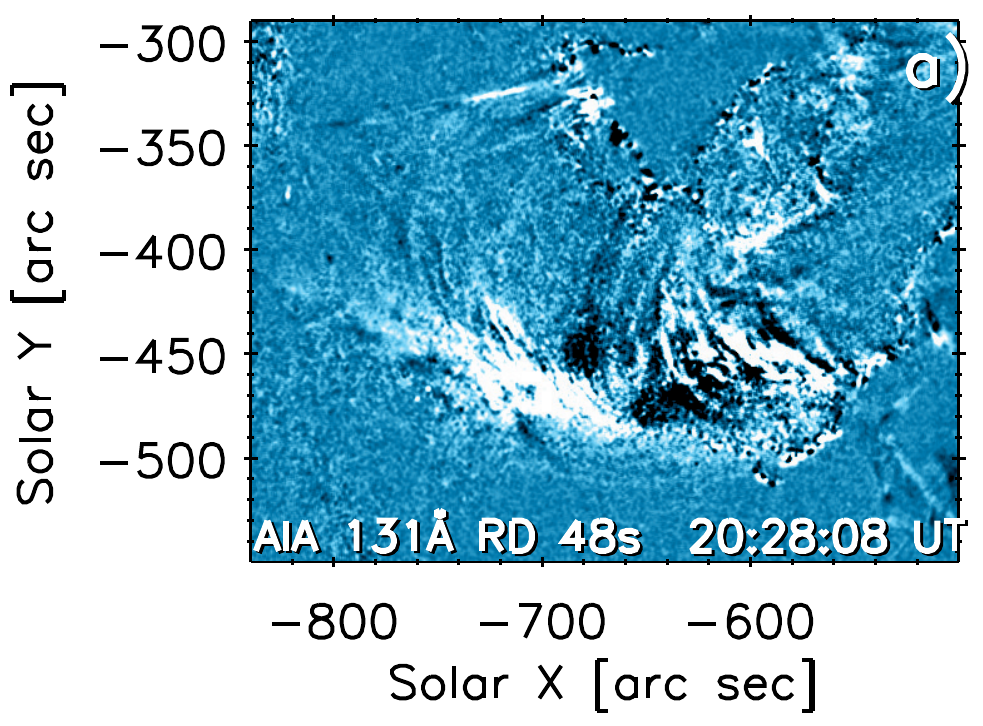}
	\includegraphics[width=4.06cm,viewport=70  0 280 205,clip]{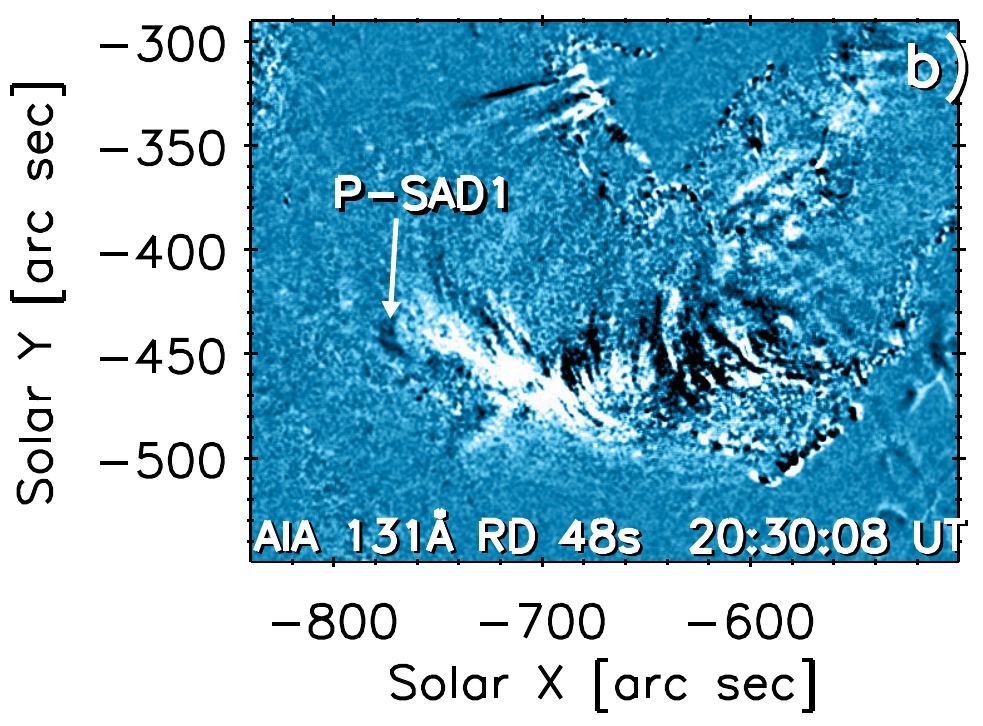}
	\includegraphics[width=4.06cm,viewport=70  0 280 205,clip]{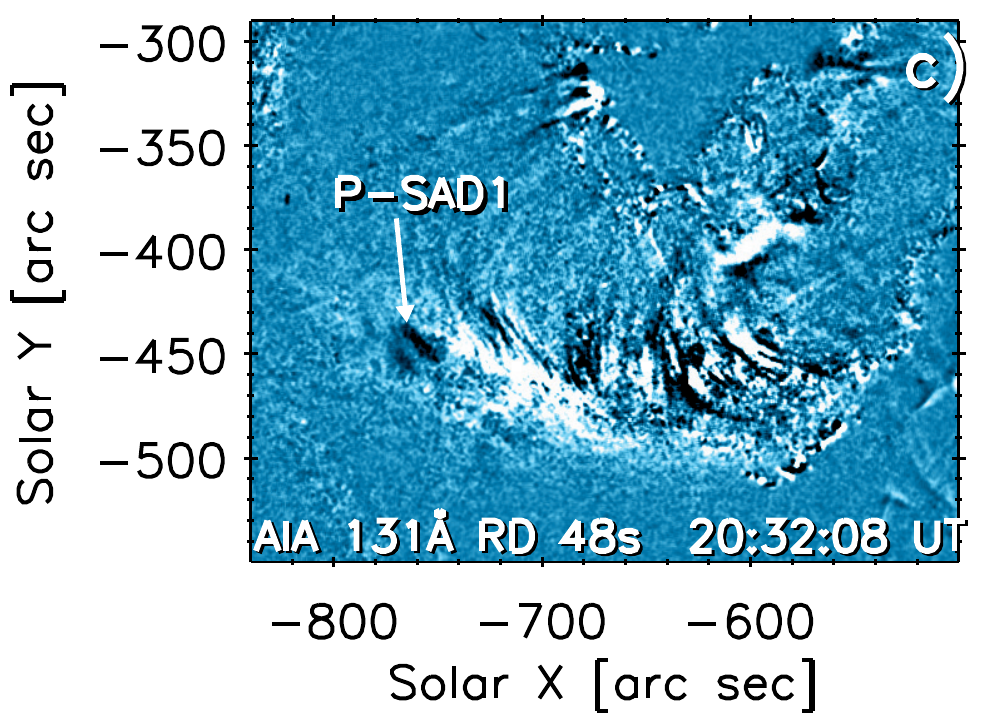}
	\includegraphics[width=4.06cm,viewport=70  0 280 205,clip]{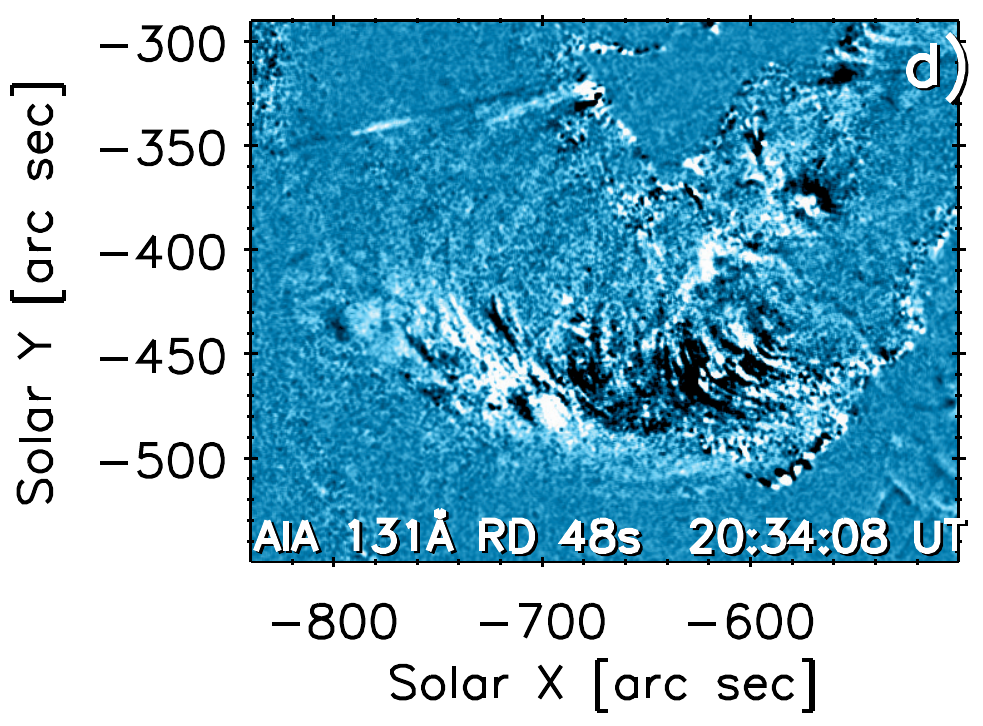}
 	\includegraphics[width=5.41cm,viewport= 0  0 280 205,clip]{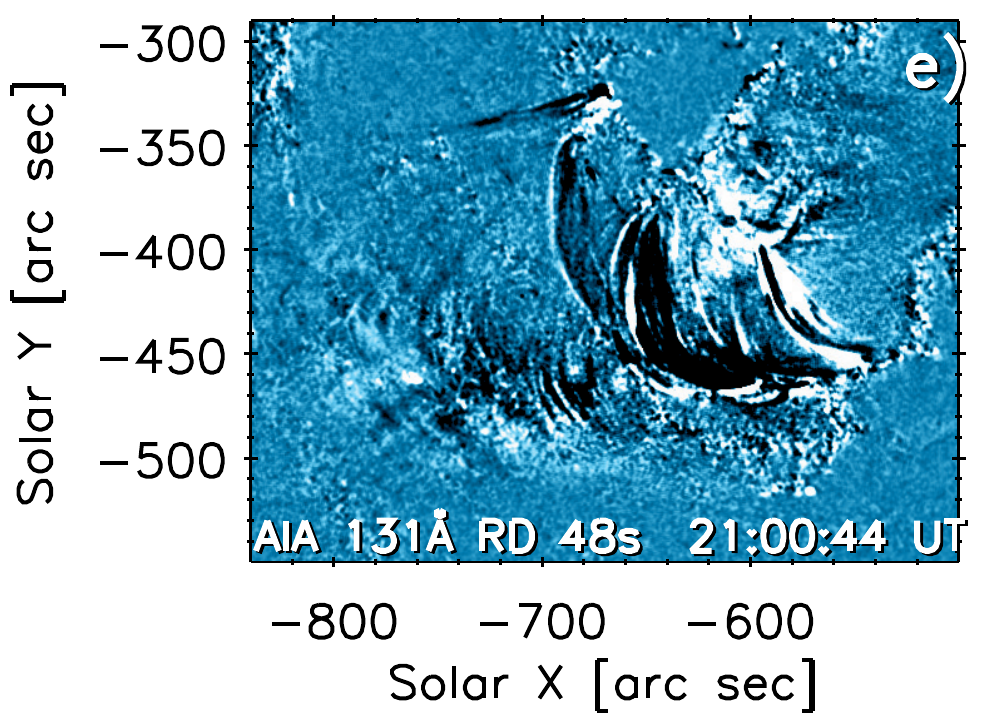}
	\includegraphics[width=4.06cm,viewport=70  0 280 205,clip]{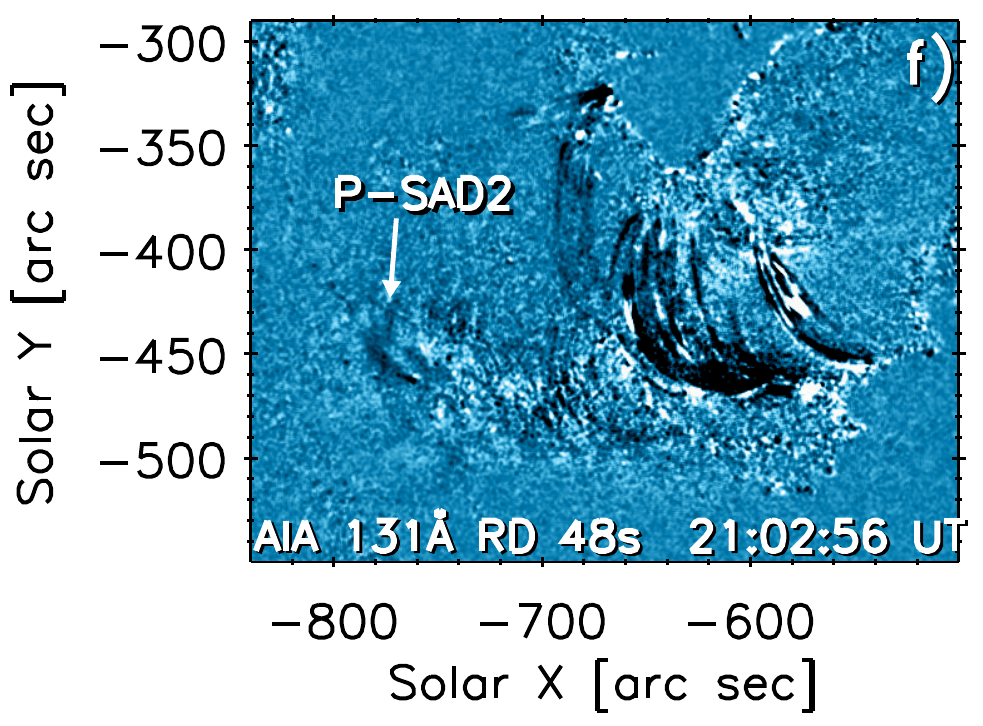}
	\includegraphics[width=4.06cm,viewport=70  0 280 205,clip]{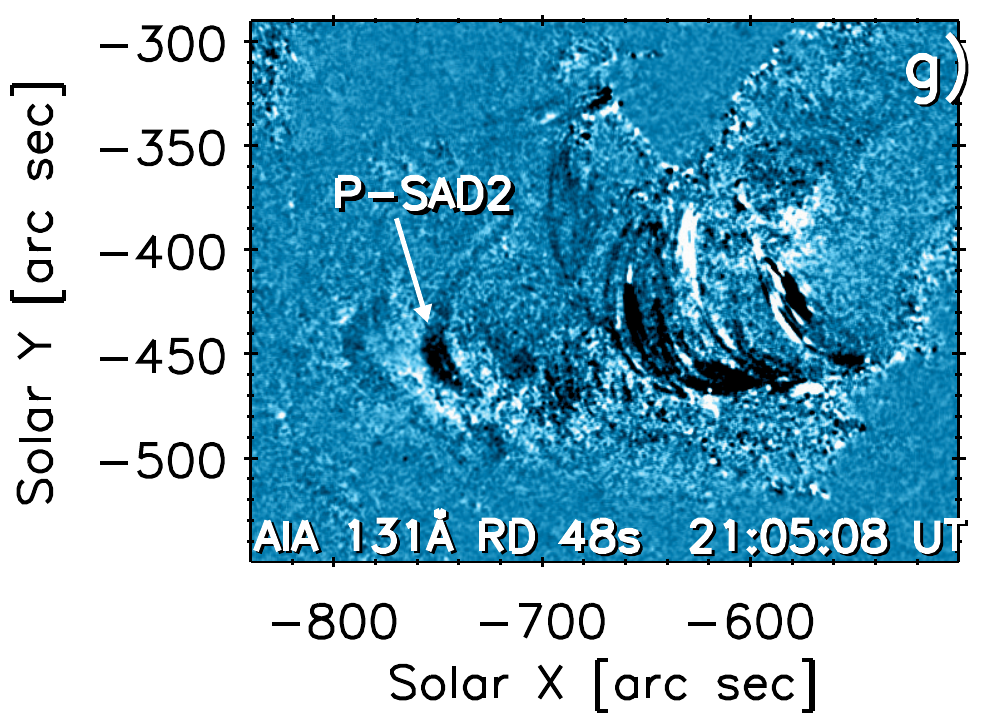}
	\includegraphics[width=4.06cm,viewport=70  0 280 205,clip]{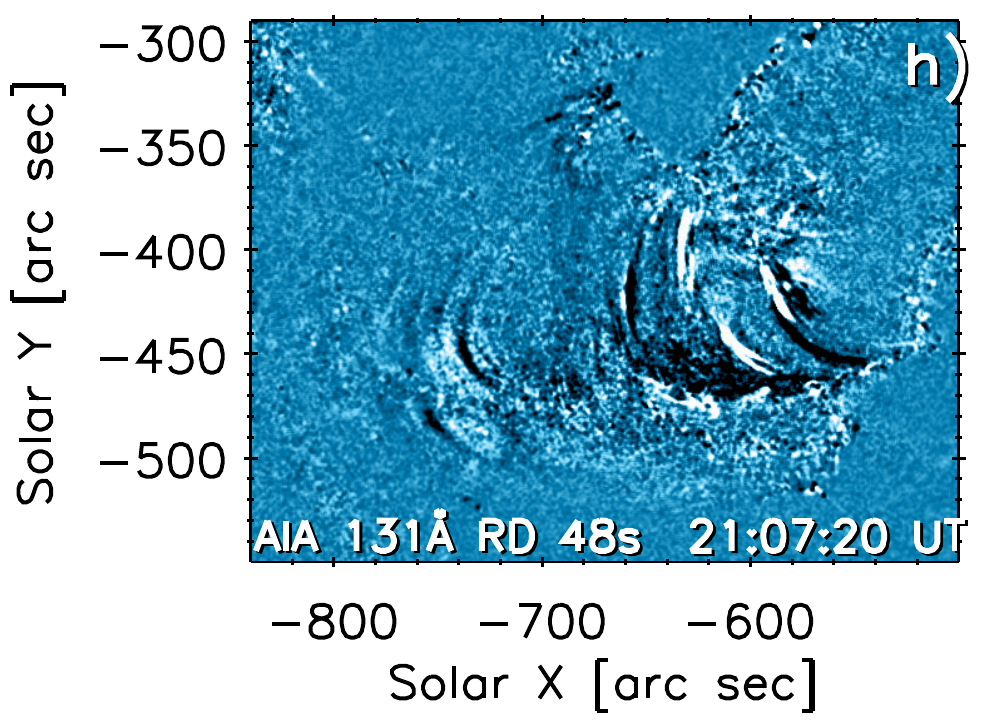}
\caption{Details of supra-arcade downflows. \textit{Top row} (a--d): SADs, including double P-SAD1, around 20:30\,UT. \textit{Bottom row} (e--h): P-SAD2 after 21:00 UT. The P-SAD2 itself is the apex part of a very faint loop-like intensity depression (panel g).}
\label{Fig:aia131_SAD}
\end{figure*}

%
%
\section{SDO/AIA data and image processing}
\label{Sect:2}

The Atmospheric Imaging Assembly \citep[AIA][]{Lemen12} onboard the Solar Dynamics Observatory images the Sun in 10 UV and EUV channels with a cadence as low as 12\,s and spatial resolution of 1.5$\arcsec$. Here, we use the data obtained in the 131\,\AA~and 304\,\AA~channels, although observations in other channels were reviewed as well. During flares, the 131\,\AA~channel is dominated by bright \ion{Fe}{21} emission, mostly in the form of flare loops, but also contains contributions from coronal \ion{Fe}{8} emission \citep[][]{ODwyer10,DelZanna13}. The 304\,\AA~channel is dominated by transition-region \ion{He}{2} Ly-$\alpha$ emission, with minor contributions from \ion{Si}{11} and \ion{Ca}{18} \citep{ODwyer10}.

To enhance weak features such as newly-reconnected flare loops and filament legs, we employ several image processing methods. The original AIA data are enhanced using the noise-adaptive fuzzy equalization method \citep[NAFE][see also Figure 2 therein]{Druckmuller13}. NAFE \footnote{\url{http://www.zam.fme.vutbr.cz/~druck/Nafe/}} calculates a linear combination of a gamma-transformed image with a fuzzy-equalized image, significantly enhancing local contrast of individual structures.

The weakest but dynamic structures at the limit of detection by AIA were also enhanced using the running difference (RD) method, where one image in a series is subtracted from a preceding one using a pre-defined temporal delay. We employ the 72\,s delay for 304\,\AA~and 48\,s delay for 131\,\AA. The shorter delay for 131\,\AA~is necessitated by presence of fast-contracting flare loops and supra-arcade downflows (see Section \ref{Sect:4}). To suppress the otherwise dominant noise, three RD images are averaged in time and additionally smoothed by a 3$\times$3 (1.8$\arcsec \times$1.8$\arcsec$) boxcar. The RD 131\,\AA~and 304\,\AA~images shown in Figures \ref{Fig:Eruption}--\ref{Fig:aia131_SAD} are saturated to $\pm$10 and $\pm$3\,DN\,s$^{-1}$, respectively, to enhance weak evolving structures.  

Finally, the off-limb 304\,\AA~images were also enhanced using the radial filter of \citet[][]{Masson14}. To keep the cadence and spatial resolution, we chose not to average or bin the individual input images. Instead, the radially-filtered AIA images are smoothed by 3$\times$3 boxcar, which suppresses noise, but not structures. We note that the radial filtering briefly fails around 21:00\,UT due to the 304\,\AA~images being saturated by another flare on the Sun.

%
%
\section{The 2012 August 31 filament eruption}
\label{Sect:3}

The spectacular long-duration eruption of a quiescent filament of 2012 August 31 (SOL2012-08-31T19:45:00) is fairly well known \citep[see the cover image of][]{Priest14}. It is one of the only two filament eruptions where the filament constituting the core of the CME (at $>$\,5\,$R_\odot$) can be traced back to the Sun \citep{Howard15b,Howard17}. The prominence mass was detected all the way out to 1\,AU~\citep{Wood16}, with some mass loss \citep{Howard15b}. The accompanying solar flare was studied by \citet{Lorincik19a,Lorincik19b}, who focused on slipping flare loops and fast kernels moving along flare ribbons \citep{Lorincik19a}, and the \textit{ar--rf} reconnection \citep{Lorincik19b}.

The overview of the eruption is shown in Figure \ref{Fig:Eruption}. The accompanying long-duration flare began at roughly 19:40\,UT and reached its peak about an hour later (panel a). Shortly after being initiated, the erupting filament shows brightenings along its inner edge (panels b, f), with bright threads extending to the chromosphere along both filament legs. The filament legs in negative and positive polarities \citep[see Figure \ref{Fig:Eruption}f, and the magnetogram in Figure 1b of][]{Lorincik19a} are denoted as FLN and FLP, respectively.

As the filament erupts, the FLP fans out and then splits into two parts, FLP1 and FLP2. The FLP1 constitutes the bulk of the leg, with multiple twisted threads rooted in the extensive positive-polarity ribbon hook PRH \citep[see Figure \ref{Fig:Eruption}b; also Figure 1b--c of][]{Lorincik19b}. The FLP2 splits from FLP as early as 19:35\,UT (see animation accompanying Figure \ref{Fig:Eruption}) as the lower part of the leg untwists during the eruption. Interestingly, the FLP2 is relatively weak in coronal AIA channels, and nearly absent in 131\,\AA~(panel f), highlighting the importance of 304\,\AA~observations. Overall, both the filament legs are noticeably thicker in 304\,\AA, which shows much more filament material (compare panels c and f). 

\subsection{Reconnection of FLN and FLP1}
\label{Sect:3.1}

As the eruption progresses, the filament legs FLN and FLP1 are stretched outwards, coming into contact for the first time at around 19:50 below the apex of the filament (Figure 1, panels b\,\&\,f and the accompanying animation). Following that, the FLP1 bends and moves northwards, towards FLN (panels c\,\&\,g). Some interaction between oppositely-oriented FLN and FLP1 is seen again at 20:10 UT high above the limb, producing bright clumps (Figure \ref{Fig:Eruption}, panels c,\&\,g) that move upwards with the eruption.

Later, at 20:26\,UT, the bulk of FLP1 runs into FLN, tearing it apart (Figure \ref{Fig:Eruption} d\,\&\,h and \ref{Fig:aia304_2028UT}). The torn FLN shows an arch-like opening above the FLP1, while the lower portions of the corresponding FLN threads disappear (Figure \ref{Fig:aia304_2028UT}b\,\&\,f). The arch-like opening subsequently stretches and moves outward with the eruption (panels c \& g), leaving behind only a weak remnant of FLN. Meanwhile, the threads of FLP1 weaken in intensity and disperse (panels d\,\&\,h). The existence of the arch-like opening above the locus of FLN and FLP1 interaction suggests that we are indeed observing leg--leg reconnection, and not just the two legs sliding past each other in projection.

\subsection{Convergence and interaction of FLN remnant and FLP2 after 21:00 UT}
\label{Sect:3.2}

The northward motion of FLP1 is followed later by northward motion of FLP2, whose emission is faint, but nevertheless visible in both the RD 304\,\AA~images (Figure \ref{Fig:Eruption}, panels e\,\&\,i) and radially filtered images (Figure \ref{Fig:aia304_2100UT}). The FLP2 is about 100$\arcsec$ wide and composed of several threads. The leading and trailing threads are most readily visible (Figure \ref{Fig:aia304_2100UT}, panels e--f). The remnant of the FLN remains visible and largely stationary until the northward-moving FLP2 converges towards it.

The interaction of FLN remnant and FLP2 starts at about 21:00\,UT (after the peak of the flare) and happens above the disk near the limb, at [$X$,\,$Y$]\,$\approx$\,[$-800\arcsec$,\,$-400\arcsec$], where the two legs approach (Figure\,\ref{Fig:aia304_2100UT}\,b\,\&\,f, and the accompanying animation). The convergence lasts during next several tens of minutes, with the convergence point seemingly moving slowly upwards. During this process, both legs gradually weaken and disappear (panels e--h). According to the time-distance plot (Figure \ref{Fig:aia304_2100UT}i), the leading thread of FLP2 is observed to last until about 21:15\,UT, while the weaker trailing thread becomes difficult to detect after about 21:35\,UT even in radially filtered images. In RD 304\,\AA~images, its signal drops below 1\,DN\,s$^{-1}$. After 21:45\,UT, well into the flare gradual phase, no trace of the filament legs is present, and only the fading flare arcade remains (see the animation of Figure \ref{Fig:Eruption}).

%
%
\section{Flare supra-arcade and Downflows}
\label{Sect:4}

The filament eruption is accompanied by a M1.2--class flare (Figure \ref{Fig:Eruption}a) according to the updated \textit{GOES}--15 data \footnote{see \url{https://hesperia.gsfc.nasa.gov/rhessidatacenter/complementary_data/goes.html}}. 
The evolution of the flare arcade in 131\,\AA~is shown in Figure \ref{Fig:aia131_arcade}a--d and the accompanying animation.

\subsection{The supra-arcade region}
\label{Sect:4.1}

The flare arcade attains maximum brightness in 131\,\AA~around 20:40\,UT (Figure \ref{Fig:aia131_arcade}c,\,f). This includes significant brightening in the supra-arcade region. This brightening and extension of the supra-arcade region during 20:00--20:25\,UT is well-visible in the time-distance plots, especially in RD 131\,\AA~(white area in Figure \ref{Fig:aia131_arcade}h), which were constructed along a horizontal cut placed at $Y$\,=\,$-450\arcsec$ (Figure \ref{Fig:aia131_arcade}e,\,g).

Figure \ref{Fig:aia131_arcade} shows that the supra-arcade region appears to be in its entirety composed of closed flare loops (see also Figure 3n--x and Figure 6a of \citet{Lorincik19b} and also compare with Figure 1 of \citet{Shen22}). Figure \ref{Fig:aia131_arcade}f shows that the supra-arcade loops are shrinking, with typical velocities of about 15\,km\,s$^{-1}$ along the cut. 

\subsection{SADs}
\label{Sect:4.2}

As soon as 20:26\,UT, that is, immediately after the supra-arcade region reaches its maximum extent, it gets permeated by dark, downflowing blobs (Figure \ref{Fig:aia131_arcade}g--h and Figure \ref{Fig:aia131_SAD}b--d). These are almost invisible in the NAFE 131\,\AA, but are apparent in the running-difference of the original data (Section \ref{Sect:2}). Their properties and dynamics (see below) mean that these are SADs \citep{Savage11,Warren11}. While the SADs occur throughout the supra-arcade region, several prominent ones (hereafter P-SADs) move along the horizontal cut at $Y$\,=\,$-450\arcsec$ shown in Figure \ref{Fig:aia131_arcade}g. The first one, P-SAD1 (Figure \ref{Fig:aia131_arcade}h), appears at around 20:28\,UT as an intensity depression shaped as loop-top (Figure \ref{Fig:aia131_SAD}b--d). It is closely followed by another one, forming a double-SAD (panel c). 
Another prominent SAD (P-SAD2) occurs after 21:00\,UT, again resembling a loop apex at first (Figure \ref{Fig:aia131_SAD}f). Few minutes later, the P-SAD2 is clearly a top part of a dark loop-like structure in the RD 131\,\AA~image (Figure \ref{Fig:aia131_SAD}g). Note that some SADs were indeed observed to be loop-shaped before \citep{Warren11,Savage11}.
Finally, a third prominent SAD (P-SAD3) occurs at about 21:35\,UT. These P-SADs move at sub-Alfv\'{e}nic velocities, which decrease over time \citep[as noted in][]{Warren11}. The P-SAD1 starts its retraction at 140\,$\pm20$\,km\,s$^{-1}$, while the P-SAD2 is faster at 300\,$\pm40$\,km\,s$^{-1}$ and later slows down to about 50\,$\pm7$\,km\,s$^{-1}$. The P-SAD3 is relatively-slower, with initial velocity of about 95\,$\pm10$\,km\,s$^{-1}$ and subsequently slowing down (Figure \ref{Fig:aia131_arcade}h).

We note that no absorption feature, such as filament plasma, is seen in the P-SADs in any AIA EUV images. Comparison of the RD and NAFE time-distance plots shows that each SAD, and P-SADs in particular, are preceded by thin supra-arcade downflowing loops in the NAFE 131\,\AA~(white arrows in Figure \ref{Fig:aia131_arcade}f). In the RD images, these thin loops are obscured by the P-SADs due to the averaging employed (Section \ref{Sect:2}). 

%
%
\section{Interpretation}
\label{Sect:5}

In all three cases, the prominent SADs1--3 occur shortly after the interaction of the conjugate filament legs, whose magnetic fields are oppositely oriented (Section \ref{Sect:3.2}): The P-SAD1 follows the interaction of FLN with FLP1, the P-SAD2 occurs after interaction of FLN remnant with leading thread of FLP2, and the P-SAD3 corresponds in time to the disappearance of the trailing FLP2 thread, which also converged to FLN remnant. These close temporal associations point to a causal relationship involving the \textit{rr--rf} reconnection, i.e., reconnection between flux rope legs, producing a new flux rope field line and a new flare loop \citep{Karlicky10,Kliem10,Aulanier19}.

Here, the \textit{rr--rf} reconnection occurs after the impulsive phase, and well into the peak and gradual phases of the flare. It lasts a long time, from formation of clumps at 20:10\,UT to about 21:35\,UT, and leads to a near-complete annihilation of the visible filament legs. The 3D extension to the Standard flare model \citep[where the leg--leg reconnection falls into the \textit{rr--rf} nomenclature,][]{Aulanier19} does not cover the gradual phase of the flare, as it stops earlier due to numerical instability during the fast eruption.
Therefore, our observations mean that the \textit{rr--rf} reconnection likely plays a substantially larger role in the evolution of solar eruptions and CMEs than the model of \citet{Aulanier19} revealed. In other models though, the leg--leg reconnection \citep{Karlicky10,Kliem10}, plays a more prominent role.
In our observations, the \textit{rr--rf} reconnection occurs during the late stage of the shrinkage of the respective negative ribbon hook \citep[denoted NRH and studied by][see Figure 2f--l therein]{Lorincik19b}. Some of the observed \textit{r}$\rightarrow$\textit{f} conversions of filament threads to flare loops reported there \citep[footpoint F5 in][]{Lorincik19b} appear to be in fact due to \textit{rr--rf}, not \textit{ar--rf} reconnections.
Finally, the occurrence of \textit{rr--rf} reconnection in shrinking hooks after the flare peak, as well as disappearance of the filament legs reaching the Sun are both consistent with the model of \citet{Jiang21}. However, quantifying the role of ar-rf and rr-rf reconnections in shrinking hooks remains a subject for future study.

We attempted to find supporting X-ray and radio evidence for reconnection during filament leg interactions after 20:26 and 21:00\,UT, but we did not succeed. The available radio emission in the MHz to GHz ranges is weak. This is perhaps not surprising, given that the reconnecting structures in AIA can be weak, indeed at the limit of detection. The HXR emission observed by \textit{RHESSI} \citep{Lin02} or \textit{Fermi} \citep{Meegan09} shows peaks, but cannot be disambiguated from other events occurring in the neighboring active regions 11560 and 11563. 

Do our observations mean that the \textit{rr--rf} reconnection is always responsible for prominent SADs? Theoretically, during the gradual phase, magnetic fields of both the flux rope and the surrounding corona become highly stretched in the radial direction. \citet{Shen22} showed that outflows from the current sheet leads to SADs, but their model did not distinguish between \textit{aa--rf} and \textit{rr--rf} reconnection geometries. Nevertheless, stronger magnetic fields in the erupting flux rope compared to the surrounding corona \citep[see Figure 11 in][]{Barczynski19} should lead to stronger magnetic tension in post-reconnection flare loops originating in the \textit{rr--rf} reconnection, and thus to their faster contraction through the supra-arcade and prominent SADs.

Finally, we reviewed the AIA observations of 15 other SAD events \citep[C to X-class, see][]{Warren11,Savage12a,Savage12b,Innes14,Hanneman14,Reeves17,Chen17,Xue20,LiZ21}, and found that about two thirds are not associated with filament eruptions. For example, the well-studied event of 2011 October 22 \citep[see, e.g.,][]{Savage12a,Hanneman14,Reeves17,Xue20,LiZ21} or the 2011 May 09 one \citep{Warren11,Lorincik21} are eruptions of hot flux ropes. Even though prominent SADs do occur there, they happen long after the hot flux rope has erupted. Direct imaging  the \textit{rr--rf} (leg--leg) reconnection in such events will require novel instruments for observing the far off-limb corona.

To summarize, our observations strongly suggest that the \textit{rr--rf} (leg--leg) reconnection can be persistent throughout the peak and gradual phases of a flare and causes strong changes in evolution of the erupting filament, and presumably the CME. In the low corona, it also leads to occurrence of prominent supra-arcade downflows (P-SADs).


\begin{acknowledgments}
The authors thank the referee for comments that helped improve the manuscript.
J.D., J.K., M.K., and A.Z. acknowledge support from the Grant No.\,20-07908S 
of the Grant Agency of the Czech Republic, as well as institutional support RVO:67985815 from the Czech Academy of Sciences. G.A. acknowledges financial support from the French national space agency (CNES), as well as from the Programme National Soleil Terre (PNST) of the CNRS/INSU also co-funded by CNES and CEA. M.D. acknowledge support from the Grant Agency of Brno University of Technology, project No. FSI-S-20-6187.
SDO data were obtained courtesy of NASA/SDO and the AIA and HMI science teams. 
\end{acknowledgments}


\end{document}